\documentclass{mn2e} 
\usepackage{epsfig}
\usepackage{times}

\newif\ifAMStwofonts

\title[Kinetic Power of radio AGN]
      {Measuring the kinetic power of AGN in the radio mode}

\author[Merloni \& Heinz] {Andrea Merloni$^{1,2}$ \& Sebastian Heinz$^{3}$\\$^{1}$Max-Planck-Institut f\"ur Astrophysik,
Karl-Schwarzschild-Strasse 1, D-85741, Garching,
Germany\\$^{2}$Max-Planck-Institut f\"ur Extraterrestrische Physik,
Giessenbachstr., D-85741, Garching, Germany\\$^{3}$Astronomy
Department, University of Wisconsin-Madison, Madison, WI 53706}

\date{}

\begin{document}

\maketitle

\label{firstpage}

\begin{abstract}
We have studied the relationship among nuclear radio and X-ray power,
 Bondi rate and the kinetic luminosity of
sub-Eddington active galactic nuclear (AGN) jets, as estimated from
the $pdV$ work done to inflate the cavities and bubbles observed in
the hot X-ray emitting atmospheres of their host galaxies and clusters.
Besides the recently discovered correlation between jet kinetic
 and Bondi power, we show that a clear correlation exists also
between Eddington-scaled kinetic power and bolometric luminosity, 
given by: $\log (L_{\rm kin}/L_{\rm Edd}) = (0.49\pm0.07) \log (L_{\rm
bol}/L_{\rm Edd}) - (0.78\pm0.36)$. The measured
 slope suggests that these objects are in a
radiatively inefficient accretion mode, and has been used to put
 stringent constraints on the properties of the accretion
 flow. Interestingly, we found no statistically significant
 correlations between Bondi power and bolometric AGN luminosity, apart
 from that induced by their common dependence on $L_{\rm kin}$, 
thus confirming the idea that most of the accretion
 power emerges from these systems in kinetic form. We have
then analyzed the relation between kinetic power and radio core
luminosity. Combining the measures
of jet power with estimators of the un-beamed radio flux of
the jet cores as, for example, the so-called 'fundamental plane' of active
black holes, we are able to determine, in a statistical sense, both the probability 
distribution of the mean jets Lorentz factor, that peaks at
$\Gamma_{\rm m}
\sim 7$, and the {\it intrinsic} 
relationship between kinetic and radio core luminosity (and thus the
 jet radiative efficiency), that we
estimate as: $\log L_{\rm
kin}=(0.81 \pm 0.11)\log L_{\rm R} + 11.9^{+4.1}_{-4.4}$, 
in good agreement with 
theoretical predictions of synchrotron jet models. 
With the aid of these findings, quantitative assessments of
kinetic feedback from supermassive black holes in the radio mode
(i.e. at low dimensionless accretion rates) will be possible based
on accurate determinations of the central engine properties alone, such
as mass, radio core and/or X-ray luminosity. As an example, we suggest
 that Sgr A$^{*}$ may follow the same correlations of radio mode AGN,
 based on its observed radiative output as well as on estimates of the
 accretion rate both at the Bondi radius and in the inner flow. If
 this is the case, the supermassive black hole in the Galactic center
 is the source of $\sim 5 \times 10^{38}$ ergs s$^{-1}$ of mechanical
 power, equivalent to about 1.5 supernovae every $10^{5}$ years.
\end{abstract}


\section{Introduction}
\label{sec:intro}
The growth of supermassive black holes (SMBH) through mass
accretion is accompanied by the release of enormous amounts of
energy which, if it is not advected directly into the hole (see
e.g. Narayan \& Yi 1995), can be either radiated away, as in Quasars
and bright Seyferts, or disposed of in kinetic form through
powerful, collimated outflows or jets, as observed, for example, in Radio
Galaxies. The feedback exerted by such a powerful energy release on
the surrounding gas and stars is imprinted into the observed
correlations between black hole mass and galaxy properties
\cite{gebhardt:00,ferrarese:00,marconi:03}, as well as into the
disturbed morphology of the hot, X-ray emitting atmospheres of
groups and clusters harboring active SMBH in their centers 
\cite{boehringer:93,fabian:00,churazov:02,birzan:04,fabian:06}.
Radiative and kinetic feedback differ not only on physical grounds, 
in terms of coupling efficiency with the
ambient gas, but also on evolutionary grounds. The luminosity dependent
density evolution parameterization of the AGN luminosity functions
\cite{hasinger:05,hopkins:07} implies a delay between the epoch of
Quasar dominance and that of more sedate Radio Galaxies
\cite{merloni:04,merloni:06h}, such that kinetic energy feedback plays an
increasingly important role in the epoch of cluster formation and
virialization.

From the theoretical point of view, the idea that AGN were responsible
for the additional heating needed to explain the cooling flow riddle
(see e.g. Fabian et al. 2001 and reference therein) in clusters of galaxies
has been put forward by many in recent years
\cite{binney:95,begelman:01,churazov:01,churazov:02,dallavecchia:04,sijacki:06,cattaneo:07}.
However, only very recently was the specific importance of
mechanical (as opposed to radiative) feedback 
for the heating of baryons within the deepest dark
matter potential wells fully acknowledged by semi-analytic
modelers of cosmological structure formation
\cite{croton:06,bower:06}.
Within these schemes, because the bright quasar population peaks at
too early times, a so-called ``radio mode'' of SMBH growth is invoked
in order to regulate both cooling flows in galaxy clusters
 and the observed sizes and colors of
the most massive galaxies \cite{springel:05,croton:06}.  

Such indirect evidence for a radio mode of AGN activity
resonates with a recent body of work aimed at understanding the
physics of low-luminosity AGN. Detailed 
multi-wavelength studies of nearby galaxies
 have revealed a clear tendency for lower
luminosity AGN to be more radio loud as the Eddington-scaled accretion
rate decreases (see Ho 2002, and references
therein; Nagar, Falcke and Wilson 2005).
Scaling relations with black hole X-ray binaries (BHXRB)
 \cite{merloni:03,falcke:04,churazov:05} 
have also helped to identify AGN analogues
of low/hard state sources, in which the radiative efficiency of the
accretion flow is low \cite{pellegrini:05a,chiaberge:05,hardcastle:07} 
and the  radio emitting jet
carries  a substantial fraction of the overall power in kinetic 
form \cite{koerding:06b}.

Nonetheless, despite such a widespread consensus on the importance of
kinetic energy feedback from AGN for galaxy formation,
  the kinetic luminosity of an AGN remains very difficult
  to estimate reliably. Recent incarnations of structure formation
models (either numerical or semi-analytic) hinge on the unknown 
efficiency with which growing black holes convert accreted rest mass into jet
power. Constraints on this quantity, such as those recently presented
in Heinz, Merloni \& Schwab (2007) are clearly vital for the robustness of
the models.

Observationally, progress has been possible recently, thanks to
 deep X-ray exposures of nearby elliptical galaxies  and clusters
 which have allowed the first direct estimates
  of $L_{\rm kin}$ by studying the cavities, bubbles and weak shocks
  generated by the radio emitting jets in the intra-cluster medium
  (ICM) \cite{birzan:04,allen:06,rafferty:06}.  Two main results
 emerged from these studies: first, it appears that AGN are
 energetically able to balance radiative losses from the ICM
 in the majority of cases \cite{rafferty:06}; second, at
 least in the case of a few nearby elliptical galaxies, there is an
 almost linear correlation between the jet power and the Bondi
 accretion rates calculated from the observed gas temperatures and
 density profiles in the nuclear regions \cite{allen:06}. The
 normalization of this relation is such that a significant fraction of
 the energy associated with matter entering the Bondi radius must be
 released in the jets.

Here we address a different, complementary issue. By collating all
available information on the nuclear (AGN) properties of the sources
for which the kinetic luminosity was estimated, we 
construct a sample of sub-Eddington accreting supermassive black holes
with unprecedented level of information on both the inner
(black hole mass) and the outer (mass supply
through the Bondi radius) boundary condition for the accretion flow,
as well as a reliable inventory of the energy output, either in
radiative or kinetic form. We demonstrate here that with the aid of 
this new set of information, strong
constraints can be placed on the accretion properties of these objects.
In so doing, we derive a new, robust estimator of the kinetic power of 
a SMBH based on its nuclear properties alone, namely its mass and 
instantaneous X-ray and radio (core) luminosity. 

The structure of the paper is as follows:
In section~\ref{sec:bol} 
we study the relationships among nuclear radio and X-ray
power, Bondi rate and the kinetic luminosity of the AGN in the sample
and show that a clear relationship exists between Eddington scaled
kinetic power and hard X-ray luminosity. We then analyze the relation
between kinetic power and radio core luminosity (\S \ref{sec:rad}). In 
section~\ref{sec:disc}, we discuss our results and present a simple
coupled accretion-outflow disc model which is capable to explain the
main features of the observed sample. Finally, we
 summarize our conclusions in \S \ref{sec:conc}.

\section{The relationship between kinetic power and nuclear
bolometric luminosity}
\label{sec:bol}

\begin{table*}
\label{tab:table}
\caption{Main properties of the sample studied}
\label{tab_1}
\begin{tabular}{lccccccc}
\hline
\hline
Object & D &Log $L_{\rm R}$ & Log $L_{\rm
  X}$ & Log $M_{{\rm BH,}\sigma}$ & Log $L_{\rm kin}$ & $P_{\rm Bondi}$
& References \\
(1) & (2) & (3) & (4) & (5) & (6)  & (7) & (8) \\
\hline
Cyg A$^{a}$ & 247 & 41.43 & 44.22 & 9.40$^{b}$ & 45.41$^{+0.19}_{-0.1}$ & 45.24$^{c}$ & 1,2,3,4,5\\
NGC 507 & 71.4 & 38.80 & $<$39.90 & 8.90 & 44.01$^{+0.16}_{-0.26}$ &
44.41$\pm 0.09$  & 6,7,8\\
NGC 1275 (Per A)$^{a}$ & 77.1 & 40.74 & 43.40 & 8.64 &
44.33$^{+0.17}_{-0.14}$ & 44.31$^{c}$ & 6,9,4,10\\
NGC 4374 (M84) & 17 & 38.43 & 40.34 & 8.80 & 42.59$^{+0.6}_{-0.5}$ & 43.69$^{+0.30}_{-0.29}$ & 6,11,8,4\\
NGC 4472 & 17 & 36.69 & 38.46 & 8.90 & 42.91$^{+0.14}_{-0.23}$ & 43.79$^{+0.25}_{-0.23}$ & 6,7,8 \\
NGC 4486 (M87)$^{a}$ & 17 & 38.88 & 40.55 & 9.48$^{b}$ &
43.44$^{+0.5}_{-0.5}$ & 44.15$^{+0.28}_{-0.40}$ & 6,12,13,4,8,14 \\
NGC 4552 (M89)& 17 & 38.23 & 39.33 & 8.57 & 42.20$^{+0.14}_{-0.21}$ &
43.37$^{+0.22}_{-0.21}$ & 6,7,8\\
NGC 4636$^{a}$ & 17 & 36.40 & $<$38.40 & 8.20 & 42.65$^{+0.11}_{-0.15}$ &
42.29$^{+0.24}_{-0.24}$ & 6,15,8,16\\
NGC 4696 & 44.9 & 39.10 & 40.26 & 8.60 & 42.89$^{+0.22}_{-0.22}$ & 43.40$^{+0.56}_{-0.55}$ & 17,4,8\\
NGC 5846 & 24.6 & 36.50 & 38.37 & 8.59 & 41.86$^{+0.18}_{-0.29}$ & 42.85$^{+0.43}_{-0.43}$& 6,7,8\\
NGC 6166 & 135.6 & 39.95 & 40.56 & 8.92 & 43.82$^{+0.5}_{-0.4}$ &
43.49$^{+0.34}_{-0.26}$ & 18,19,4,8\\
IC 4374 & 94.5 &  40.27 & 41.37 & 8.57 & 43.30$^{+0.36}_{-0.26}$ & 44.37$^{c}$ & 20,4\\
UGC 9799 & 152 & 40.55 & 41.89 & 8.58 & 44.18$^{+0.36}_{-0.28}$ & 43.92$^{c}$
& 21,22,4\\
3C 218 (Hydra A) & 242 & 40.91 & 42.17 & 8.96 & 44.63$^{+0.16}_{-0.10}$ &
44.90$^{c}$ & 23,24,4\\
3C 388 & 416 & 40.69 & 41.69 & 9.18 & 44.30$^{+0.38}_{-0.30}$ & 44.80$^{c}$ & 25,4\\
\hline
\end{tabular}
\vskip 0.3cm Notes: $^{a}$ Objects for which a measure of the kinetic
jet power from modelling of either jets and radio lobe emission or shocks
exists; $^{b}$ dynamical mass measurements; $^{c}$ Bondi power
extrapolated from measures of gas temperature and density outside the
Bondi radius, assuming a $r^{-1}$ density profile.

Col. (1): Name of the object. Col. (2): Adopted distance in
Mpc. Col. (3): Logarithm of nuclear (core)
luminosity at 5 GHz. Col. (4): Logarithm of the
intrinsic rest-frame luminosity in the 2-10 keV band. Col. (5) Logarithm of
the black hole mass as derived from $M-\sigma$ relation. Col. (6)
Logarithm of the Kinetic Luminosity; when multiple estimates
available, the logarithmic mean is used. Col. (7) Logarithm of the
Bondi Power as defined in eq. (\ref{eq:pbondi})

REFERENCES: (1) Sambruna et al. (1999); (2) Young et al. (2002); (3)
Tadhunter et al. (2003); (4) Rafferty et al. (2006); (5) Carilli \&
Barthel (1996); (6) Nagar et al. (2005); (6) De Ruiter et al. (1986);
(7) Pellegrini (2005b); (8) Allen et al. (2006); (9) Allen et
al. (2001); (10) Fabian et al. (2002); (11) Terashima et al. (2002);
(12) Di Matteo et al. (2003); (13) Macchetto et al. (1997); (14)
Bicknell and Begelman (1996); (15) Loewenstein et al. (2001); (16) Jones et
al. (2007); (17) Taylor et al. (2006); (18) Giovannini et al. (1998);
(19) Di Matteo et al. (2001); (20) Johnstone et al. (2005); (21) Zhao
et al. (1993); (22) Blanton et al. (2003); (23) Zirbel \& Baum (1995);
(24); Simpson et al. (1996); (25) Evans et al. (2006)
\end{table*}

We have collected from the literature data on the
nuclear properties of
AGN with published measurements of the jet kinetic power, as estimated 
from the  $pdV$ work done to inflate the cavities and bubbles observed in
the hot X-ray emitting atmospheres of their host galaxies and clusters. 
We begin by  considering the samples of Allen et al. (2006) (9
sources) and Rafferty et al. (2006) (33 sources). By taking into
account common objects, there are 38 AGN with kinetic power estimated
in such a way. Out of those, we consider only the ones  
for which black hole mass could be estimated (either through
the $M-\sigma$ relation or via direct dynamical measurements), that
amount to 21 objects. Finally, we search in the literature for 
available measures of the nuclear luminosity in the radio (at 5 GHz) and in
the 2-10 keV band. Only 6/21 do not have such information, and we
further notice that 4 out of these 6 are at a distance equal to, or larger
than, that of the most distant object in our final sample (with the
exception of Cygnus A). Therefore, there are only two objects (NGC
708 and ESO 349-010, both from the Rafferty et al. 2006 sample) which
have a measure of $L_{\rm kin}$, an estimate of the black hole mass
and are close enough ($z<0.055$) to qualify as sample members, but
which have been dropped due to the lack of X-ray and radio nuclear
luminosities. Such a high level of ``completeness'' of the sample
should guarantee us against substantial selection bias due, for example, to
beaming effects in the radio band (see section~\ref{sec:rad}
below)\footnote{It is of course important to notice that the overall 
selection criteria of the sample discussed here are indeed
heterogeneous. In particular, the role played by the 
requirement of having a bright, hot
X-ray emitting atmosphere against which detect bubbles and cavities is
very hard to assess, as surface brightness effects and different
exposure times of the original observations should all 
contribute to the measurability of kinetic power.}.

Our selection provided us with
a sample of 15 objects, 2 of which have only upper limits to their
2-10 keV nuclear X-ray luminosity. In four cases, estimates of the jet
power are also available based on detailed modelling of either radio emission
in the jets and lobes (Cyg A, Carilli \& Barthel 1996; M87, Bicknell
\& Begelman 1996; Per A, Fabian et al. 2002), or shock in the IGM
induced by the expanding cocoons (NGC 4636, Jones et al. 2007). 
In another handful of objects (M84, M87, NGC 4696, NGC 6166) both
Allen et al. (2006) and Rafferty et al. (2006) report (independent) measures of the
jet kinetic power, which, incidentally, differ on average by almost one
order of magnitude. In all these case when more than one measurement
of the kinetic 
power was found, we have used their logarithmic average for our
study, and increased the uncertainty accordingly. These are 
reported in Table~\ref{tab:table}, where a summary of the sample
adopted is presented.  For all these objects we
define the Bondi power as 
\begin{equation}
\label{eq:pbondi}
P_{\rm Bondi} \equiv 0.1 \dot M_{\rm Bondi} c^2,
\end{equation}
where the Bondi accretion rate is calculated from measures of gas
temperatures and density under the assumption of spherical symmetry
and negligible angular momentum (see Allen et al. 2006, and references
therein). For the nine AGN studied in Allen et al. (2006) the Bondi
rate is calculated extrapolating the observed inner density profile
of the X-ray emitting gas down to the Bondi radius. For the others,
more luminous (and more distant) objects from the Rafferty et
al. (2006) sample, the {\it Chandra} resolution corresponds to a size
several orders of magnitude larger than the true Bondi radius. For
these objects, the Bondi power is estimated by assuming a $r^{-1}$
inner density profile. This inevitably implies very large
uncertainties on $P_{\rm Bondi}$, that we have taken 
into account by increasing the nominal error bars associated
with the measurements found in the literature.

As mentioned in the introduction, our sample provides 
us with information on both the inner
and the outer boundary condition for the accretion flow,
as well as a reliable inventory of the energy output, both in
radiative and kinetic form. In principle, if all these objects were to
be described by the same physical model for the accretion flow
(coupled to the jet/outflow), one should expect simple scaling
relationships between the mass supply at the outer boundary (Bondi
power), and the energy emitted in kinetic or radiative form from the
accretion flow ($L_{\rm kin}$ and $L_{\rm X}$, respectively).

We first perform a partial correlation analysis in order to test
which, if any, correlation between these quantities is statistically
significant in our sample. We choose the partial Kendall's $\tau$ correlation test
for censored data sets
\cite{akritas:96}. The results of the correlation analysis are shown in
Table~\ref{tab:part}. The correlations between the kinetic power and
both Bondi power and X-ray luminosities are significant at more than
3-sigma level. Interestingly, the correlation between $L_{\rm
X}$ and $P_{\rm Bondi}$ does not appear to be statistically
significant once the common dependence of these two quantities on
$L_{\rm kin}$ is accounted for. This is consistent with previous
studies \cite{pellegrini:05a,soria:06} which failed to detect any
clear correlation between $L_{\rm X}$ and $P_{\rm Bondi}$, 
and might indicate that the X-ray
luminosity, and thus the accretion power released radiatively, is
not sensitive to the outer boundary condition, while it depends more
strongly on the mechanical output of the flow. This is indeed
expected, for example in adiabatic inflow-outflow (ADIOS) 
models \cite{blandford:99,blandford:04}, but is also true for more
general disk-wind models provided that a negligible fraction of the 
binding energy of the accreting gas is converted directly into
radiation \cite{merloni:02,kuncic:04}, 
as we will show in  more detail in 
section~\ref{sec:disc}.

\begin{table*}
\caption{Results of partial correlation analysis}
\label{tab:part}
\begin{tabular}{lllccc}
\hline
\hline
\multicolumn{3}{c}{Variables} &  
\multicolumn {3}{c}{Correlation}\\
X & Y & Z & $\tau$& $\sigma_K$ & P$_{\rm null}$ \\
(1) & (2) &(3) &(4) & (5) &(6) \\
\hline
Log $L_{\rm X}$ & Log $L_{\rm Kin}$ & Log $P_{\rm Bondi}$ & 0.49 & 0.1818 &  $7.0 \times 10^{-3}$\\
Log $P_{\rm Bondi}$ & Log $L_{\rm Kin}$ & Log $L_{\rm X}$ & 0.5872 &
0.2815 &  $3.7\times 10^{-2}$ \\
Log $L_{\rm X}$ & Log $P_{\rm Bondi}$ & Log $L_{\rm Kin}$ & 0.1289 & 0.1422 &   $0.3647$\\
\hline
\end{tabular}
\vskip 0.3cm NOTE: Col. (1): Variable X. Col. (2): Variable Y. Col (3):
Variable Z. Correlation between variables X and Y is studied, taking into
account the mutual correlation of X and Y with Z. Col. (4)-(6):
Results of partial
correlation analysis, giving the partial Kendall's $\tau$ correlation
coefficient, the square root of the calculated variance $\sigma_K$, and the
associated probability $P_{\rm null}$ for accepting the null hypothesis
that there is no correlation between X and Y.
\end{table*}

On physical grounds, one might expect that both the black hole mass
and the accretion rate should determine the power channeled through
the jet, thus enforcing the need to perform multivariate
statistical analysis on any multi-wavelength database of AGN, a point
already discussed in MHD03. However, we notice here that the sample
in question spans only a very limited range of black hole
masses\footnote{This is likely the result of a specific selection
bias, as the method used to estimate the average jet kinetic power can
only be used effectively for radio galaxies at the center of bright
X-ray emitting atmospheres, as those of clusters of galaxies (X-ray
surface brightness selection). This in turn tend to select the more
massive elliptical galaxies, with large central black holes.}. 
Attempts to quantitatively account for the mass dependence of
the kinetic luminosity based on such a sample were made, and yielded negative
results. Therefore in what follows we limit ourselves to a simple
bivariate correlation
analysis between Eddington-scaled quantities.  
Future, more detailed, studies of
the physical connection between jet kinetic power and black hole
accretion processes will greatly benefit from samples with much wider
distributions of black hole masses. 

\begin{figure}
\psfig{figure=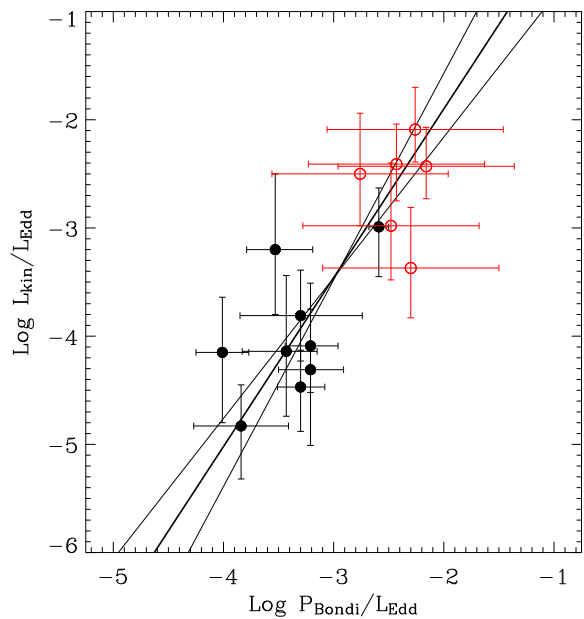,width=0.5\textwidth}
\caption{Eddington-scaled kinetic power vs. Bondi power
 for the AGN in the sample. 
The thick solid line shows the best fit linear regression, with one-sigma
uncertainties (thin solid lines). Red open circles with larger error bars
indicate those objects for which the Bondi radius is much smaller than
the resolution and Bondi power is calculated by
extrapolating inwards the observed density assuming a $r^{-1}$ profile.}
\label{fig:lkinlbondi}
\end{figure}

Figure~\ref{fig:lkinlbondi} shows the relationship
between Bondi power and kinetic luminosity, both 
in units of the Eddington luminosity, $L_{\rm Edd}=4 \pi M_{\rm BH}
m_{\rm p} c / \sigma_{\rm T}$, where $m_{\rm p}$ is the mass of a proton and
$\sigma_{\rm T}$ is the Thomson cross-section.
The best fit linear regression slope, estimated via a symmetric
algorithm that takes into account errors on both variables (see MHD03)
gives:
\begin{equation}
\label{eq:lkbondi}
\log (L_{\rm kin}/L_{\rm Edd}) = (1.6^{+0.4}_{-0.3}) \log (P_{\rm
Bondi}/L_{\rm Edd}) + (1.2^{+1.0}_{-0.8})
\end{equation}
consistent, within the 1-$\sigma$ uncertainty,
with the best fit slope found by Allen et al. (2006) (1.3$^{+0.45}_{-0.27}$).

We have then studied the relationship between kinetic power
and bolometric (radiated) luminosity. The bolometric correction for
objects of such a low power is not well known. In general terms, it
has been argued that low
luminosity AGN in nearby galaxies display a spectral energy
distribution markedly different from classical quasars, lacking 
clear signs of thermal UV emission usually associated to standard,
optically thick accretion discs (see e.g. Ho 2005, and references
therein; for a contrasting view, Maoz 2007).  
The bolometric luminosity is probably dominated by the hard X-ray
emission, but quantitative assessments of the bolometric corrections
for low luminosity AGN based on well defined samples are still
missing. Here, for the sake of simplicity, we adopt a common 2-10 keV 
bolometric correction factor of 5 for all objects in our sample, and
define  $\lambda_{\rm X}=5L_{\rm X}/L_{\rm Edd}$.
 
The best fit linear regression slope, estimated via a symmetric
algorithm that takes into account errors on both variables\footnote{We
assume here that the errors on $\lambda_{\rm X}$ are dominated by the
statistical uncertainties on the black hole mass determinations, which
we estimate to be of the order of $\sim$0.2 dex from the intrinsic scatter in
the M-$\sigma$ relation, see Tremaine et al. (2002).}, gives the following result:
\begin{equation}
\label{eq:lklambda}
\log (L_{\rm kin}/L_{\rm Edd}) = (0.49\pm0.07) \log \lambda_{\rm X} - (0.78\pm0.36)
\end{equation}
with an intrinsic scatter of about 0.39 dex.

\begin{figure}
\psfig{figure=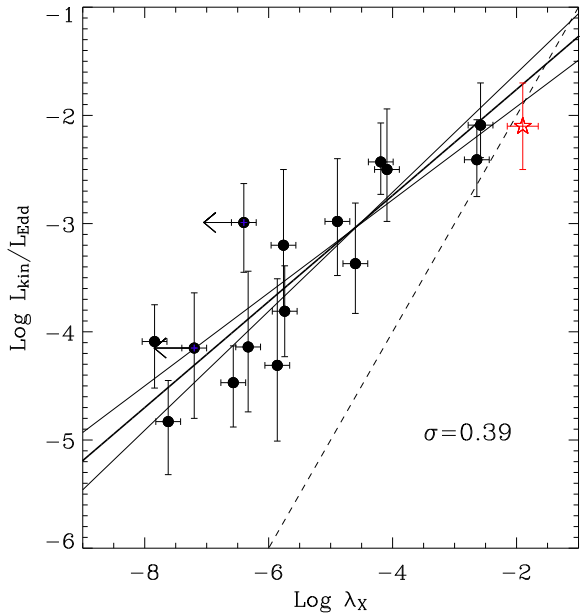,width=0.5\textwidth}
\caption{Eddington-scaled kinetic power vs. bolometric nuclear
luminosity for the AGN in the sample (black solid circles) and for the
BHXRB Cyg X-1 (empty red star). The solid line shows the best fit linear regression, while
the dashed line is the $L_{\rm kin}=5L_{\rm X}$ relation shown as a
term of comparison.}
\label{fig:lkinledd}
\end{figure}


The best fit relationship together with the data is shown in
Figure~\ref{fig:lkinledd}, and is
{\it inconsistent} with a linear relationship at more than 4-$\sigma$ level.
Thus, not only are these SMBH more radio loud the
lower their accretion rate (a fact already suggested by many previous
works both on nearby AGN and X-ray binaries: Ho \& Ulvestad 2001; Gallo, Fender \&
Pooley 2003; MHD03; Panessa et al. 2007), but
their "kinetic loudness", i.e. the ratio of kinetic to
bolometric powers (akin to a jet to accretion efficiency) 
is also a decreasing function of the dimensionless accretion rate onto the
black hole.
Even more interestingly, the best-fit slope coincides with the
theoretical expectations of radiatively inefficient, "jet-dominated"
accretion modes \cite{fender:03}. We will come back
to this point in greater detail in section~\ref{sec:disc}.

As discussed in the introduction, many recent works have explored 
scaling relationships between black
holes of stellar mass and supermassive ones by studying
multidimensional correlation among different observables
(MHD03; McHardy et al. 2006), supporting the notion that AGN are
indeed scaled-up galactic black holes. However, physical models for the
disc-jet coupling in BHXRB based on the observed correlations between
radio and X-ray luminosity \cite{fender:03,koerding:06b} all face a
large uncertainty due to the lack of reliable measurements of the jet
kinetic power. In this context, it is interesting to include 
in Figure~\ref{fig:lkinledd} 
the only GBH for which a  measurement of the kinetic output has been
made, Cyg X-1 \cite{gallo:05}, which turns out to be consistent with
the relationship derived from the AGN sample. Clearly, systematic
efforts to estimate kinetic power of BHXRB in the low/hard state are
needed in order to assess their similarity with radio mode AGN.

\section{The relationship between kinetic power and radio core luminosity}
\label{sec:rad}

If the above correlation (\ref{eq:lklambda}) directly reveals fundamental physical
properties of jet-producing AGN of low power, it still shows a
non-negligible intrinsic scatter. 
On the other hand, one should expect a more
direct relationship between the nuclear radio core emission and the
larger scale kinetic power, as both originate from the jet.
All theoretical models for AGN flat-spectrum compact jet cores
\cite{blandford:79,falcke:96,heinz:03} {\it predict} a dependence of the
radio luminosity on the jet power in the form $L_{\rm R} \propto
L_{\rm kin}^{17/12}$. The current sample provides by far the best
opportunity to test these predictions.

\begin{figure}
\psfig{figure=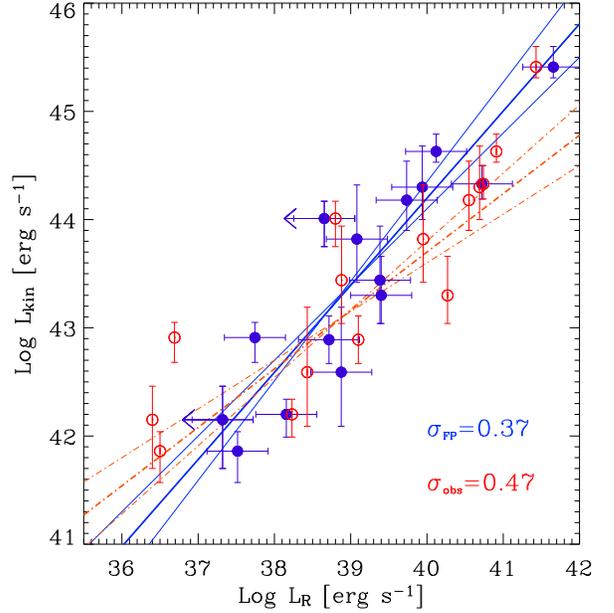,width=0.5\textwidth}
\caption{Kinetic power, as measured from estimates of $pV$ work by the
jet on its surrounding hot, X-ray emitting, atmosphere versus radio
core luminosity of the jets at 5 GHz. Red open circles show the
observed radio luminosity (not-corrected for relativistic beaming),
while the blue filled circles refer to the radio luminosity estimated
through the ``fundamental plane'' relation derived from the KFC
sample, and corrected for relativistic
beaming, with a mean Lorentz factor of the jets of $\Gamma_{\rm
m}\sim 7$ (see text for details). Thick (and thin) solid lines show
the best fit to this latter relationship, with 1-$\sigma$ uncertainty;
dashed lines show the best fit linear relation when the observed radio
luminosity is considered. The value of the intrinsic scatter for the
two cases is reported.}
\label{fig:pboth}
\end{figure}

A Kendall's tau correlation test reveals that the kinetic power is
correlated with the {\it observed} radio core luminosity $L_{\rm
R,obs}$, with $P_{\rm null}= 9.2\times 10^{-5}$ (see
the empty circles in Figure~\ref{fig:pboth}). 
We have fitted the data with a linear relationship: 
$\log L_{\rm kin} = A_{\rm obs} + B_{\rm obs} \log L_{\rm R,obs}$,
once again making use of a symmetric regression algorithm that takes
into account errors on both variables. We obtain 
$A_{\rm obs}= (22.1 \pm 3.5)$, $B_{\rm obs}=(0.54 \pm 0.09)$, 
with a large intrinsic scatter of $\sigma=0.47$. 

Such a correlation, however, must be at some level biased by relativistic Doppler
boosting of the radio emission in the relativistic jets. 
An alternative way to proceed would be 
to use {\it indirect} estimators of
the nuclear radio core luminosity which are less affected by
relativistic beaming \cite{heinz:05}, as, for example, the
multivariate relation between BH mass, radio core and hard X-ray
luminosity, the so-called `fundamental plane' (FP) of active black
holes (MHD03). Recent analysis of this
correlation (Heinz \& Merloni 2004; K\"ording, Falcke \& Corbel 2006
[KFC06]; Merloni et al. 2006) have shown that both Doppler boosting
and sample selection play a crucial role in the exact determination of
the intrinsic correlation coefficients of the FP, which also need to be
accounted for. In the Appendix, we discuss in detail
a possible way to overcome such a bias with the aid 
of a Monte Carlo simulation of the samples used to derive the FP relation.
That study allows us to estimate statistically the intrinsic (un-boosted) radio core
luminosity of the AGN jets as a function of their (mean) Lorentz
factor, $\Gamma_{\rm m}$ in a way that can be approximated by the
following expression:
\begin{eqnarray}
\label{eq:fp_corr}
\log L_{\rm R,FP}&=&(1-0.14 \log \Gamma_{\rm m})[\xi_{\rm RX} \log L_{\rm
  X}+\xi_{\rm RM} \log M_{\rm BH} ] \nonumber \\ && + c_{\rm R}(\Gamma_{\rm
  m}),
\end{eqnarray}
where $ L_{\rm R,FP}$ is the intrinsic (un-boosted) radio core
luminosity of the jet at 5 GHz, $L_{\rm X}$ the nuclear 2-10 keV
intrinsic (un-absorbed) luminosity and $\Gamma_{\rm m}$ the mean
Lorentz factor of the jets.

In what follows, we will adopt the specific version of the FP relation
derived from a
sample of low luminosity AGN only, i.e. free from the bias introduced
by the inclusion of bright, radiatively efficient AGN or QSOs (see discussion in
KFC06). For that, the correlation coefficients are 
 $\xi_{\rm RX}=0.71$, $\xi_{\rm RM}=0.62$, slightly different (but
only at the 1-$\sigma$ level) from those found in MHD03.

\begin{figure}
\centering
\psfig{figure=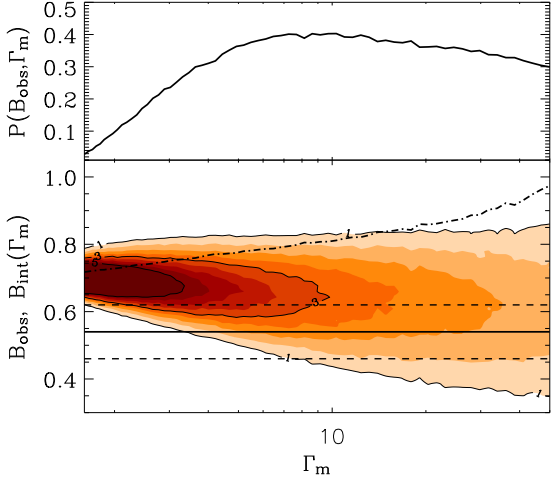,height=7cm}
\caption{Bottom
  panel: probability density of the observed  slopes
  of the $L_{\rm Kin}$ - $L_{\rm R,obs}$ relation (colored
contours): For each value of the mean jets Lorentz factor,
  $\Gamma_{\rm m}$, the intrinsic correlation has a
  slope $B_{\rm int}$ determined from the best estimate of the beaming
correction of the fundamental plane relation used as estimator of the
intrinsic radio core luminosity (see text for details). 
The dot-dashed line in the lower panel shows
  this $B_{\rm int}$, while the solid line shows the observed slope of
  the real data set with its uncertainty (dashed lines). The upper
  panels show then the integrated probability of observing $0.45<B_{\rm
  obs}<0.63$, given $B_{\rm int}$ and $\Gamma_{\rm m}$.}
\label{fig:bobs}
\end{figure}

Given Eq.~(\ref{eq:fp_corr}), assuming 
a distribution of Lorentz factors for the AGN jets (or its mean,
provided that the distribution is not too broad), we 
determine the ``true'' relationship between the  Kinetic luminosity and
radio core luminosity by fitting the 
15 data points in our sample with the linear relationship
\begin{equation}
\label{eq:lkin_int}
\log L_{\rm kin} = A_{\rm int}(\Gamma_{\rm m}) + B_{\rm
  int}(\Gamma_{\rm m}) \log L_{\rm R,FP}.
\end{equation}
The fitted values for the intrinsic slope, $B_{\rm int}$, 
as a function of $\Gamma_{\rm m}$, are shown as a dot-dashed line in
the bottom panel of Figure~\ref{fig:bobs}. 
From this we can see that the higher the mean Lorentz factor of the
jets,  the steeper must the
intrinsic correlation between kinetic power and jet core luminosity be,
and the larger the discrepancy with the measured slope, $B_{\rm
obs}=0.54 \pm 0.09$ 
(solid lines in Figure~\ref{fig:bobs}), obtained using simply the
observed radio core luminosity, without any attempt to correct for
relativistic beaming.  
In fact, such a discrepancy between the intrinsic and the observed
slopes of the $L_{\rm R}$ - $L_{\rm kin}$ relation is indeed 
expected if the 15 sources of
our sample harbor relativistic jet randomly oriented with respect to
the line of sight\footnote{It is worth noting here that the sample has 	
no a priori selection against beamed objects (indeed, 3C 84, the radio source
at the core of NGC 1275, is most likely beamed, Kirchbaum et
al. 1992).}. 

In order to show this quantitatively, we have simulated ($10^4$ times)
the {\it observed} sample, assuming an underlying
relationship between intrinsic radio core luminosity and kinetic
luminosity given by Eq.~(\ref{eq:lkin_int}). In order to do that we
have assumed distribution distances, black hole masses radio and X-ray
core flux limits that closely resemble the observed ones. 
We then Doppler-boost the intrinsic radio core luminosity picking
$\Gamma$ from a normal 
distribution with mean $\Gamma_{\rm m}$ and
variance $\sigma_{\Gamma}=0.1\Gamma_{\rm m}$. 
Fits of the $L_{\rm kin}$-$L_{\rm R,obs}$ correlation for the $10^4$
simulated samples result in a distribution of slopes $B_{\rm obs}$ as a
function of $\Gamma_{\rm m}$ (shaded areas in the lower panels
figure~\ref{fig:bobs}). We can now assess the probability of observing 
$B_{\rm obs}=(0.54 \pm 0.09)$ for
any value of $\Gamma_{\rm m}$, by simply integrating these (properly
normalized) distributions in the range 0.54$\pm$0.09. The result of
such an integration is
shown in the upper panel of Figure~\ref{fig:bobs}.

The distribution of the simulated sources in the $L_{\rm R,obs}/L_{\rm
R}$ vs. $L_{\rm kin}$ plane shows tantalizing evidence of a
discrepancy with the observed sample, with a slight deficit of
luminous, de-boosted (i.e. seen at large angles with respect to the
line of sight) objects. Statistically, we found that this discrepancy
mainly effects the normalization of the intrinsic $L_{\rm
kin}$-$L_{\rm R}$ relation, rather than its slope. The reason for
such a discrepancy is not clear at this point, but we believe it may
signal a problem in the modellization of the selection criteria of the
sample, rather than a problem with the FP scaling. This is by no means
surprising, in particular given the lack of any realistic constraint
on the distribution and selection functions of the kinetic power
measurements.

From this Monte Carlo test, we can reach the following conclusions:
\begin{itemize}
\item[(i)]{For any given $\Gamma_{\rm m}$, the beaming-corrected FP
relation (\ref{eq:fp_corr}) can be used to derive the {\it intrinsic}
radio core luminosity of the jets, $L_{\rm R,FP}$;} 
\item[(ii)]{The correlation between $L_{\rm kin}$ and $L_{\rm R,FP}$ in
our sample has a slope $B_{\rm int}(\Gamma_{\rm m})$, that is an increasing
function of the mean jet Lorentz factor;}
\item[(iii)]{The relation between
$L_{\rm kin}$ and {\it observed} radio core luminosity 
is flattened with respect to the intrinsic one (i.e. $B_{\rm
obs}<B_{\rm int}$) because of the
unaccounted-for Doppler boosting in the small sample at hand. We have
corrected for such an effect statistically: 
Figure~\ref{fig:bobs} shows that a fully consistent interpretation of
the data can be given for a broad range of
mean Lorentz factors, with a broad probability distribution that peaks at 
$\Gamma_{\rm m} \simeq 7$, corresponding to the following intrinsic
correlation:
\begin{equation}  
\label{eq:lklr}
\log L_{\rm kin}= (0.81\pm0.11) \log L_{\rm R} + 11.9^{+4.1}_{-4.4}
\end{equation}
Reassuringly, the
probability distribution of mean Lorentz factors, although broad,
clearly excludes low values of $\Gamma_{\rm m}$. More detailed studies
of the relativistic speeds of Blazars emitting regions give similar
results (see e.g. Cohen et al. 2007).}
\end{itemize}

Interestingly, the relationship between kinetic power and radio luminosity
gets tighter if one uses the FP relation as an estimator of the
un-boosted radio flux, as shown in Figure~\ref{fig:pboth}: the
intrinsic scatter is reduced from 0.47 dex to 0.37 dex.
This is expected if this steeper correlation reflects
more directly the intrinsic relation between jet kinetic and radiative
power, and lends support to our statistical method to correct for the
bias induced by relativistic beaming.

Finally, we note that the slope of the correlation between $L_{\rm R,FP}$ and
$L_{\rm kin}$ is consistent with the predictions of synchrotron models
for the flat spectrum jet cores according to which \cite{heinz:03}
$L_{\rm R} \propto L_{\rm kin}^{(17+8\alpha_{\nu})/12}
M^{-\alpha_{\nu}}$, where $\alpha_{\nu}$ is the radio spectral index. 
For flat spectrum cores, like the ones we are
considering here, $\alpha_{\nu}\sim 0$ and the logarithmic slope of
the $L_{\rm kin}-L_{\rm R}$ correlation is expected to be $\sim
12/17=0.71$, within 1-$\sigma$ from our estimate.

The intrinsic relation between radio and kinetic luminosity of the
jets can also be used to determine the total 
radiative efficiency of the jet cores, $\eta_{\rm jet}$, the ratio of
radiated to kinetic luminosity. 
This will of course depend on the
maximal frequency, $\nu_{\rm max}$, up to which the flat radio core 
spectrum extends. Assuming a constant spectral index (i.e. neglecting
spectral steepening at $\nu<\nu_{\rm max}$) we have:
\begin{equation}
\log \eta_{\rm jet} = (\alpha_{\nu}+1) \log (\nu_{\rm max}/5 {\rm
GHz}) + 0.19 \log L_{\rm R} -11.9
\end{equation} 
Many nearby low-luminosity radio galaxies observed with the {\it HST} show 
unresolved optical nuclei the flux of which clearly correlates
with the radio one \cite{chiaberge:99}, suggesting that in those
object the jet emission extends up to the optical band, and the
corresponding radiative efficiency can be as high as 10-20\%. This
result is not substantially modified if we allow at least a non
negligible fraction of the X-ray core luminosity to originate in the
jet, as recently argued on the basis of either X-ray spectroscopy
alone \cite{evans:06} or on model
fitting of multiwavelength SED \cite{wu:07}. This is because the
available constraints on broad band SED of low-luminosity radio AGN
(mostly FR I) based on
radio-to-optical-to-X-ray indexes (see e.g. Hardcastle \& Worrall
2000; Heinz 2004; Balmaverde, Capetti \& Grandi 2006) 
show that the spectra must be concave everywhere
between the radio and X-ray bands, and this alone ensures 
that the radiative output of the
jet cores is dominated by the emission at the peak frequency,
$\nu_{\rm max}$, thus justifying the
use of the above equation.

\section{Discussion}
\label{sec:disc}

\subsection{Testing accretion flow models}
\label{sec:model}
In section~\ref{sec:bol} we have shown that there is a significant
correlation between kinetic power and Bondi power, as well as between
X-ray (bolometric) luminosity and kinetic power, while the Bondi power
and the X-ray luminosity are not correlated. Here we try to interpret
this fact, as well as the observed slopes of the statistically
significant correlations within simple models for sub-Eddington
accretion flows. 

The idea that low-luminosity AGN accrete in radiatively inefficient
fashion was first put forward by Rees et al. (1982) and Fabian \&
Rees (1995), while efforts were made at the same time to
accommodate advection dominated accretion flows (ADAF; Narayan \& Yi
1995) within the general framework of accretion theory. Very early on
it was also recognized that such kind of adiabatic (i.e. non radiative)
accretion flows are prone to the production of strong outflows, so
that in fact only a tiny fraction of the gas entering the black hole
sphere of influence finally makes its way onto the black hole
\cite{blandford:99,blandford:04}. While the observational evidence for
low radiative efficiency in many nearby galactic nuclei has grown
significantly in recent years
 (see e.g. Ho 2002; MHD03; Pellegrini 2005; Chiaberge et al. 2005)
there remains substantial uncertainty on the true dynamical state of
such flows.

Here we want to exploit the unique possibility offered by this sample
with  its comprehensive information on both inner and  outer 
boundary conditions for the accretion flow,
as well as its reliable inventory of the energy output. 
We start from a very general accretion
disc model coupled to a disk wind such as those studied in detail in
Kuncic \& Bicknell (2004). There the interested reader will find an
analytic description of the most general MHD disk accretion,
specifically addressing the relationship between radial and vertical
mean field transport of mass, momentum and energy.

In what follows, we will use as a starting point Eq.~(84) of Kuncic \&
Bicknell (2004) for the rate of energy released by the viscous torques
within the disc as:
\begin{equation}
\label{eq:fd}
Q^{+} = \frac{3GM_{\rm BH}\dot M_{\rm out}}{8\pi
r^3}\zeta_{\rm K}(r) -
\frac{3\Omega_{\rm K}}{8 \pi r^2}\int_{r_{\rm in}}^{r_{\rm out}}
\frac{\dot M_{w}(r') v_{\rm K}(r')}{2} dr'
\end{equation}
where $\dot M_{\rm out}$ is the accretion rate at the outer boundary,
$\zeta_{\rm K}=1-(r_{\rm in}/r)^{1/2}$, $\Omega_{\rm K}$ is the
Keplerian angular velocity $\Omega_{\rm K}=v_{\rm K}/r=\sqrt{GM/r^3}$,
$\dot M_{w}(r)$ is the mass loss rate in the outflow, 
and we have neglected any additional torque at the disc inner
boundary, as well as all additional terms due to vertical Poynting
flux and re-irradiation of the disc from outside, for the sake of simplicity.

The first term on the right hand side of Eq.~(\ref{eq:fd}) represents
the binding energy flux associated with the net mass flux, and is
familiar from standard accretion disc theory \cite{shakura:73}, while
the second term is related to the flux of kinetic and gravitational
energy in the outflow.

We adopt here the simplest phenomenological model for the accretion
rate through the disc:
\begin{equation}
\dot M_a(r)= \left\{
        \begin{array}{ll}
        \dot M_{\rm out}, & \hbox{$r \geq r_{\rm out}$} \\
        \dot M_{\rm out} \left(\frac{r}{r_{\rm out}}\right)^{p}, & \hbox{$r < r_{\rm out}$}   \\
        \end{array}\right.\;
\end{equation}
with $0 \le p < 1$ \cite{blandford:99}, 
where $r_{\rm out}$ represents a critical radius
beyond which mass outflow is negligible. Correspondingly, mass
conservation implies that the mass outflow as a function of radius can
be expressed as $\dot M_{w}=\dot M_{\rm out} - \dot M_{\rm a}= \dot M_{\rm
out}[1-(r/r_{\rm out})^{p}]$. With such a simple ansatz,
the kinetic energy properties of the outflow closely resemble those of
the adiabatic ADIOS solutions.

By integrating Eq.~(\ref{eq:fd}) from the inner radius, $r_{\rm in}$
to $r_{\rm out}$, we obtain for the power dissipated within the disc:
\begin{eqnarray}
\label{eq:ld}
P_{\rm d}&=&\int_{r_{\rm in}}^{r_{\rm out}}\!\!\!\!\! 4 \pi r Q^{+} dr=
P_{\rm a}-L_{\rm mech} \nonumber \\
&=& \frac{GM\dot M_{\rm out}}{2r_{\rm
in}}\left[\frac{\xi^p}{(1-p)}-\frac{3}{(2p+1)(1-p)}\xi\right. \nonumber \\
&& \left.+\frac{2}{(2p+1)}\xi^{p+3/2}\right]
\end{eqnarray}
where we have introduced the parameter $\xi\equiv \frac{r_{\rm
in}}{r_{\rm out}}$. In the above expression, $P_{\rm a}$ represents
the total power released by the accretion process, while $L_{\rm
mech}$ is the mechanical energy carried by the outflow.
The latter is given by
\begin{eqnarray}
\label{eq:mech}
L_{\rm mech} &=& v_{\infty}^2 \int_{r_{\rm in}}^{r_{\rm out}}\! \frac{3
\Omega_{\rm K}}{2r}\int_{r_{\rm in}}^{r}v_{\rm K}(r')r'\frac{d\dot
M_{\rm a}}{dr'}dr' \nonumber \\
&=&\frac{GM\dot M_{\rm out}v_{\infty}^2}{2r_{\rm in}}\frac{2p}{(1-p)}
\left[\xi^p-\frac{3}{(2p+1)}\xi\right. \nonumber \\
&& \left.+\frac{2(1-p)}{(2p+1)}\xi^{p+3/2}\right]
\end{eqnarray}
where $v_{\infty}$ is the ratio of the terminal wind speed to the
Keplerian velocity, here assumed to be independent of $r$. 

The radiative power depends on micro-physical properties of the disc
(capability of heating electrons as opposed ions),
on its vertical structure and on the radiative processes
involved. Radiatively efficient discs, as those discussed in Kuncic \&
Bicknell (2004), would obviously have $L_{\rm
bol}=P_{\rm d}$, and we can already guess that this option is not
allowed by the observed correlations. 
Indeed, in the limit of very large $r_{\rm out}$, $\xi \ll 1$, we have that
both the mechanical luminosity of the disc, $L_{\rm mech}$ and $P_{\rm
d}$ scale as the accretion rate through the disc in its inner part
$\dot M_{\rm a} \sim \dot M_{\rm out} \xi^p$, so that
$L_{\rm mech}\propto L_{\rm bol}$, rather than  the observed 
$L_{\rm mech}\propto L_{\rm bol}^{0.5}$ (see
section~\ref{sec:bol}). The 
only way in which this can be altered is by considering the effect
of magnetic fields in the accretion flow. As it was proposed by
Merloni \& Fabian (2002) and discussed also in Kuncic \& Bicknell
(2007), strong magnetic field threading a thin disc may remove energy
and angular momentum (and very little mass) without radiating
efficiently, and in so doing generating powerful, Poynting dominated
outflows. However, the specific scaling between kinetic and radiative
power in this case depends in a complicated way on the dynamo properties of the
disc, and is not yet as fully understood as the adiabatic (ADIOS)
solutions we discuss here.

This suggests a simple phenomenological scaling for the radiative output of
mass-losing low luminosity discs, that 
can be written as $\log (L_{\rm bol}/L_{\rm
Edd}) = C + \gamma \log (P_{\rm d}/L_{\rm Edd})$. 
The normalization
$C$ is related to the value of the dimensionless accretion rate above
which radiatively inefficient accretion ceases to be a viable solution
due to the increase of disc density and cooling rate. 
The value of the parameter $\gamma$, instead, depends on the
radiative cooling properties of the accreting gas, and within our
approach has to be fixed by comparison with observations.

To fully characterize the model, and make it suitable to be
tested against the available data, one still needs to specify the
relationship between $r_{\rm out}$ and $\dot M_{\rm out} = \dot M_{\rm
Bondi}$. In adiabatic
solutions with outflows, the radius beyond which radiative losses are
significant and the adiabatic self-similar solution breaks down
will depend on the outer supply of mass and angular momentum and on
the cooling properties of the outer disc \cite{blandford:04}. In our case,
we simply assume:
\begin{equation}
\log \xi = A + B \log(\dot M_{\rm out}/\dot M_{\rm Edd})
\end{equation}
and use the data to put limits on $A$ and $B$.

Our accretion/outflow model has the desired
properties of simplicity and generality, which make it suitable to be
tested against the dataset at hand. Indeed, variations of the indexes
$p$ and $\gamma$ allow us to span the full range of dynamical
accretion models from pure ADAF ($p=0$ and $\gamma \simeq 2$) to ADIOS
($1>p>0$ and $\gamma \simeq 2$) to radiatively efficient, wind
dominated discs ($1>p>0$ and $\gamma \simeq 1$) down to standard,
radiatively efficient, conservative thin discs ($p=0$ and
$\gamma=1$). The parameters $A$ and $B$, in turn, give us an handle on
the large-scale geometrical properties of the flows, with $C$
representing an overall normalization of the radiative efficiency of
the system. The sixth and final parameter of the model, $v_{\infty}$,
turns out to be strongly degenerate with the normalization parameters
$A$ and $C$. Given the amount of information in the data, we have
chosen to keep its value fixed, either to a value of 1
(terminal velocity equal to the local Keplerian speed) or 2 (terminal velocity
twice the local Keplerian speed), and perform two separate fits. Detailed dynamical modelling of the jets and outflows at
different luminosities and black hole masses will be needed in order
to constrain its value reliably.

\begin{figure}
\centering
\psfig{figure=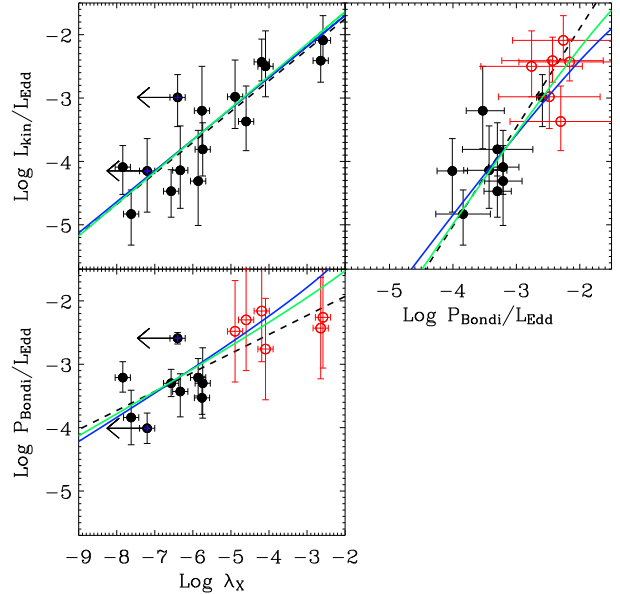,width=0.5\textwidth}
\vspace{-0.5cm}
\caption{The best fit accretion/outflow models (blue and green solid
line for $v_{\infty}=2$ and $1$, respectively) are
  plotted alongside the simple best fit linear relations among
Eddington-scaled kinetic power, Bondi power and bolometric luminosity
(see text for details).}
\label{fig:adios}
\end{figure}

In summary, our model has just five free
parameters: $p$, $\gamma$, $C$, $A$ and $B$. We have fitted this model
to the observed relationships between kinetic power and Bondi power,
and between kinetic power and bolometric luminosity.
In Figure~\ref{fig:adios} we show the best
fit to the data for the two independent correlations 
with the model described above (in the case $v_{\infty}=2$, blue solid
line and $v_{\infty}=2$, green solid line).
Our simple illustrative model provides an excellent description of the
data both in the case of $v_{\infty}=1$ and $v_{\infty}=2$. The
best-fit values for the model parameters (with their 90\%
confidence errors, calculated by letting each single parameter vary
while fixing all others to their best fit value) are given by:
$p=0.85^{+0.11}_{-0.16}$, $A = 0.13 \pm 0.30$ $B = 0.63
\pm 0.10$ and $\gamma = 1.98 \pm 0.06$ and $C=2.27 \pm 0.24$ in the
case $v_{\infty}=2$, and $p=0.92^{+0.11}_{-0.16}$, $A = 0.1 \pm 0.19$ $B = 0.45
\pm 0.10$ and $\gamma = 2.04 \pm 0.06$ and $C=2.00 \pm 0.22$ in the
case $v_{\infty}=1$. 

A crucial prediction of the model is the dependency of
$r_{\rm out}$ on the outer accretion rate. According to our fit, we
have, in units of the Schwarzschild radius $r_{\rm S}=2GM/c^2$, 
\begin{equation}
\frac{r_{\rm out}}{r_{\rm S}}\approx 6 \times 10^2 \left(\frac{\dot
    m_{\rm out}}{10^{-4}}\right)^{-0.63},
\end{equation}
where $\dot m_{\rm out}=\dot M_{\rm out}/\dot M_{\rm Edd}$. For the
objects in our sample, this is reassuringly smaller than either the
inferred Bondi radius and the radius beyond which radiative losses are
significant and adiabatic self-similar outflow dominated solutions are
not viable. In the sample under study, the two brightest objects (Cyg
A and Hydra A) have
$\dot m_{\rm out}\approx 10^{-2.5}$ and therefore outer outflow
radii of just a few tens of Schwarzschild radii. They may therefore
qualify as ``intermediate'' objects in a transition between
radiatively efficient (hidden QSO-like) and inefficient AGN. In this case, the
self-similar assumption becomes poorer, and clear signs of the
presence of an outer standard disc may be expected \cite{ogle:97,sambruna:00}. 

As we have discussed above, a clear distinction exists between those
parameters that describe the nature of the outer boundary
conditions of the flow and solely determine slope and normalization of
the $L_{\rm Kin}$-$P_{\rm Bondi}$ relation ($A$, $B$ and $p$) and
those that instead describe the nature and radiative properties 
of the inner accretion flow and determine the slope of the $L_{\rm X}$
- $L_{\rm Kin}$ relation ($\gamma$ and $p$). This might explain why we
found no statistically significant correlation between radiative output
and Bondi power. 

It is worth stressing here that the above model represents a very
schematic description of the physical properties of a low-luminosity
accretion flow and the accompanying powerful outflow. 
As mentioned above, a full exploration of all possible
theoretical alternatives for such flows should also take into account
dynamically important magnetic fields and their role in the jet
production mechanisms. The reason we have chosen not to do so here is
merely the quality of the data available, which would not have allowed
to put meaningful constraints on the parameters of such complicated
MHD disc-outflow solutions. 
What we have instead shown is that future, better samples like that assembled here 
promise to be extremely
useful for constraining the complex physical properties
 of accretion at the lowest rates.

\subsection{The kinetic power output of Sgr A*}
The quiescent radio and X-ray luminosities of the $\sim 4\times 10^{6}
M_{\odot}$ nuclear black hole
at the center of the Milky Way, 
Sgr A$^{*}$ ($\log L_{\rm R}\simeq 32.5$ and $\log L_{\rm
X}\simeq 33.3$) would be both consistent with a Bondi accretion rate  
of about $10^{-6}$ $M_{\odot}/$yr, in line with the {\it Chandra}
measurements of the inner hot gas properties of the galactic center 
\cite{baganoff:01,baganoff:03}.
If the same model outlined in section~\ref{sec:model} 
applied to Sgr A$^{*}$, it would predict
an outer radius of the outflow of about $r_{\rm out}\approx 2.5 \times
10^{3}$ Schwarzschild radii ($\sim 0.025$'' at the distance of the
galactic center) and an inner accretion rate $\dot M_{\rm a}(r_{\rm
in})=\dot M_{\rm out} \xi^{p} \simeq 4 \times 10^{-9}$
$M_{\odot}/$yr, also consistent with the gas density in the inner
accretion flow implied by linear polarization measurements
\cite{aitken:00,bower:03,bower:05}. 
Altogether, 
this would indicate that 
the SMBH at the galactic center is the source of about $5\times
10^{38}$ ergs s$^{-1}$ of mechanical power (similar to what estimated
within jet model fits to Sgr A$^{*}$ SED, see e.g. Falcke \& Biermann
1999, Falcke \& Markoff 2000), or about 1.5 supernovae every 10$^5$ years.
Such a mechanical power input into the galactic center could play a
significant role in the production of the TeV $\gamma$-rays recently
observed by the HESS (High Energy Stereoscopic System) collaboration
\cite{aharonian:04,atoyan:04}, as well as in the heating of the hot
($> 8$ keV) diffused plasma detected by {\it Chandra} \cite{muno:04}.

\subsection{The effect of variability on the observed correlations}
In section~\ref{sec:rad}, we have concentrated our attention on the
assessment of the effects of relativistic beaming on the measured
slope of the $L_{\rm kin}$-$L_{\rm R}$ correlation. However, 
one should expect a second source of scatter in any correlation
between kinetic power and nuclear luminosity (in any waveband). 
The large scale power is an average over the typical age of the
cavities and bubbles observed in the X-ray atmosphere of the galaxies
in our sample. Such an age may be assumed to be of the order of either
the buoyant
rise time, or the sound crossing time, or the refill time of the radio lobes
\cite{birzan:04}. These estimates typically differ by about a factor
of 2, and, for the sample in question, lie in the range $t_{\rm
age} \sim 10^7 - 10^8$ years.
On the other hand, the core power is variable on time scale much
shorter than that. The bias introduced by this fact can be quantified
as follows.

First of all, we notice that AGN X-ray lightcurves, where most of the
accretion power emerges, show a characteristic rms-flux relation which
implies they have a formally non-linear, exponential form
\cite{uttley:05}, and the luminosity follows a log-normal distribution
\cite{nipoti:05}. Let us, for the sake of simplicity, define as $L(t)$
the {\it instantaneous} luminosity of the AGN, be it bolometric, 
kinetic or radio, under the implicit assumption that very close to the
central engine jet and accretion are strongly coupled, so that all of
them follow a log-normal distribution. Thus, $\log L$ is normally
distributed with mean
$\mu_l$ and variance $\sigma_l^2$. 
The measured kinetic power $L_{\rm kin}$ discussed in this work being
a long-term time average, it should be determined by the mean of the
log-normally distributed $L$,
i.e. $\langle L \rangle=\mu=\exp{(\mu_l^2+\sigma_l^2/2)}$. Because 
the log-normal distribution is positively skewed, the mode $m$ of the 
distribution of
$L$, i.e. the most likely value of a measurement of it, $L_{\rm obs}$,
is {\it smaller} than the mean: $L_{\rm
obs}=m=\exp{(\mu_l-\sigma_l^2)}$. 
Therefore the most
likely value of the ratio of the mean to the observed luminosity is
given by (Nipoti \& Binney 2005; Uttley et al. 2005):
\begin{equation}
\label{eq:rms}
\frac{\langle L \rangle}{L_{\rm obs}}=e^{\frac{3}{2}\sigma_l^2}=(\sigma_{\rm rms}^2 +
1)^{3/2},
\end{equation}
where we have introduced the rms (fractional) variability of the
observed lightcurve $\sigma_{\rm rms}^2=\exp{\sigma_l^2}-1$, 
an easily measurable quantity. The higher the rms variability
of a lightcurve, the higher is the probability that an instantaneous
 measurement of the luminosity yields a value smaller than $\langle L \rangle$
\cite{nipoti:05}, and also the higher the most likely ratio between
the two values. 

Before discussing what are the appropriate values of $\sigma_{\rm
rms}^2$ to be used in Eq.~(\ref{eq:rms}), we also note that the observed
slope of the $\langle L \rangle$-$L_{\rm obs}$ correlation may deviate from unity
for a log-normal distribution, thus skewing any observed correlation
between mean and instantaneous power, like those we have discussed so
far. It is easy to show that from  Eq.~(\ref{eq:rms}) we obtain:
\begin{equation}
\frac{\partial \log \langle L \rangle}{\partial \log L_{\rm obs}}=1+\frac{3}{2} \frac{\sigma_{\rm
rms}^2}{(1+\sigma_{\rm
rms}^2)}\frac{\partial \log \sigma_{\rm
rms}^2}{\partial \log L_{\rm obs}}.
\label{eq:rms-slope}
\end{equation}
Log-normal AGN variability may thus skew the observed relation between
jet average kinetic power and instantaneous core luminosity much in
the same way relativistic beaming does.
Inspection of Eqs.~(\ref{eq:rms}) and (\ref{eq:rms-slope}) suggests
that the measured slopes of any correlation of the (average) kinetic
vs, core power will be substantially 
affected by variability if, and only if, both
$\sigma_{\rm rms}^2$ and $\partial \log \sigma_{\rm rms}^2 /\partial
\log L_{\rm obs}$ are at least of the order of unity.

Unfortunately, very little is known observationally about the
variability amplitude of AGN on very long timescales, 
especially for AGN of low luminosity as those considered here.
Brighter AGN (Seyferts), on shorter timescales (1-10 years), have indeed rms
variability amplitudes that rise steeply with decreasing luminosity,
from about $\sigma_{\rm rms}^2 \approx 10^{-2}$ for $L_{\rm X} \approx
10^{44}$ up to $\sigma_{\rm rms}^2 \approx 10^{-1}$ for $L_{\rm X} \approx
10^{42}$ \cite{nandra:97,markowitz:01}. However, no
evidence is yet found of such a trend continuing down to lower
luminosities. On the contrary, suggestions have been made that
$\sigma_{\rm rms}^2$ may flattens out at lower $L_{\rm X}$
\cite{papadakis:04,paolillo:04} at values of a few times 10$^{-1}$ for
the typical X-ray luminosity of the objects in our sample. 
This would imply that the observed correlations between large scale
and core powers are not skewed by variability bias.

However, we should also consider the possibility that, if measured on
the $t_{\rm age}$ timescale, rather than on years or decades, AGN
variability amplitude can be much higher, perhaps exceeding unity
(Nipoti \& Binney 2005). Specifically for the case of radio galaxies,
various indirect pieces of evidence of intermittency have been
presented, such as the ripples and
shock waves detected in the X-ray emitting atmosphere of M87
\cite{forman:05}, or the number vs. size counts of small radio galaxies
\cite{reynolds:97}, both obviously relevant for our current discussion.

All AGN Power Spectral Densities
(PSD: the variability power $P(\nu)$ at frequency $\nu$, or timescale $1/\nu$)
are best fitted by a power-law of index $-1$ ($P(\nu) \propto \nu^{-1}$)
on long timescales, which breaks to a steeper slope on timescales
shorter than a break timescale, which itself appear to be correlated
with the central black hole mass and accretion rate ($t_{\rm
break}\propto M_{\rm BH}^2/L_{\rm bol}$, McHardy et al. 2006).
Indeed, if the PSD of AGN at very low frequencies, $\nu\sim 1/t_{\rm
age}$, are a seamless
extension of the flicker noise observed on days-years timescales, 
then their rms variability amplitude
could be very large: $\sigma_{\rm rms}^2 \simeq 1-10$. 

If we define the duty cycle of $L(t)$ as $\delta\equiv
\langle L
\rangle^2/\langle L^2 \rangle=(1+\sigma_{\rm rms}^2)^{-1}$
\cite{ciotti:01}, then large rms amplitudes would imply very bursty  
lightcurves, with very short duty cycles $\delta \sim 10^{-1} - 10^{-2}$.  
Within this picture, the measured $L_{\rm kin}$ would be dominated by
very short, ``quasar-like phases'', rather than by quasi-continuous radio
mode AGN activity. 

Although this possibility cannot be easily ruled
out, we feel it is currently disfavored for two main reasons: first
of all, it would imply a typical QSO lifetime for these objects of
$t_{\rm Q}\la 10^6$ yrs, much less than current estimates
\cite{martini:04}; secondly, if the variability properties of low
luminosity AGN are scaled up versions of those of low/hard state GBH,
as it is the case for for bright (high/soft state) objects
\cite{mchardy:06},
then we should expect a second, lower frequency
break to white noise slope ($P(\nu)=$ const.) 
in the PSD \cite{uttley:05}. Such a break should be present
 at timescales $t_{\rm flat}$  between 10-100 times
longer than $t_{\rm break}$, but still much smaller that $t_{\rm
age}$. In this case, $\sigma_{\rm rms}^2 \simeq N [2+\ln
(t_{\rm flat}/t_{\rm break})]\simeq 6 N$, where $N$ is the frequency
independent amplitude of  $\nu\times P(\nu)$ in the flicker noise
part of the PSD, 
which has been measured for many AGN and GBH to be of the order
of a few times $10^{-2}$ \cite{papadakis:04,done:05}. Thus, at most, we should expect
$\sigma_{\rm rms}^2 \sim$ a few times 10$^{-1}$ also for low
luminosity AGN on very long timescales, implying a much higher duty
cycle (not much smaller than unity).

To summarize, the effects of the AGN intermittency on the observed
relations between time-averaged kinetic power and instantaneous core
luminosities will be comparable to that of beaming if AGN have
very bursty lightcurve (duty cycles less than 10\%) on
timescales comparable to age of the bubbles over which the kinetic
power is estimated, {\it and} the overall burstiness of the lightcurve
(its measured rms variability) increases with decreasing luminosity.
Although this possibility cannot be ruled out, there are reasons
to believe that the rms variability of the low luminosity AGN in our sample is
smaller than unity. If this is the case, the difference between instantaneous
and average power for these objects is of the order of (or smaller than)
the systematic uncertainties on the power itself, and the observed
correlations between $L_{\rm kin}$ and nuclear luminosities should not be
strongly affected by variability. More work is needed to
investigate this issue, which is beyond the scope of this paper.

\section{Conclusions}
\label{sec:conc}
We have presented a statistical analysis of a 
sample of 15 nearby AGN for which the average kinetic
power has been estimated from the study of the cavities and bubbles
produced by the jets in the IGM. In particular, rather than focusing
on the relationship between the kinetic power and the IGM physical
state, as was done before, we have tried to derive the relationship
between jet kinetic power and nuclear properties of the AGN,
specifically in terms of their black hole masses, 2-10 keV and 5 GHz
radio core luminosities. 

Following our analysis, we reach the following conclusions:
\begin{itemize}
\item{A clear relationship exists
between Eddington-scaled kinetic power and bolometric luminosity, 
given by: $\log (L_{\rm kin}/L_{\rm Edd}) = (0.49\pm0.07) \log (L_{\rm
bol}/L_{\rm Edd}) - (0.78\pm0.36)$. The measured
 slope suggests that these objects are in a
radiatively inefficient accretion mode.}
\item{We confirm previous claims of a correlation between Bondi power
    (i.e. accretion rate at the Bondi radius) and Kinetic luminosity
    of the jets. Interestingly, there is no statistically significant
    correlation between Bondi power and bolometric luminosity apart
    from that induced by their common dependence on $L_{\rm kin}$.}
\item{The observed correlations are in very good
    agreement with theoretical predictions of adiabatic accretion
    models with strong outflows. We have also shown that meaningful
    constraints on some specific physical properties of such models can
    be placed by fitting the observed data set.} 
\item{The available measures of the average jet
power provide a very useful tool to assess, in a statistical sense,
both the jet radiative efficiency and the effects of relativistic
beaming on the observed AGN jet population. Combining information
on the kinetic jet power with estimators of the un-beamed radio flux of
a jet core (as, for example, the so-called fundamental plane of active
black holes, MHD03), we are able to determine {\it both} the probability 
distribution of the mean jets Lorentz factor, that peaks at $\Gamma
\sim 7$, {\it and} the intrinsic 
relationship between kinetic and radio core luminosity, $\log L_{\rm
kin}=(0.81 \pm 0.11)\log L_{\rm R} + 11.9^{+4.1}_{-4.4}$  
which is in good agreement with 
theoretical predictions of synchrotron jet models:
$L_{\rm kin} \propto L_{\rm R}^{12/(17+8\alpha_{\nu})}$, where
$\alpha_{\nu}$ is the radio spectral index of the jet core.}
\item{The total radiative efficiency of the jet can be expressed as a
function of the observed 5 GHz luminosity ($L_{\rm R}$) and spectral
index, $\alpha_{\nu}$ as:
$\eta_{\rm jet}\simeq 3.2 \times 10^{-5} (L_{\rm R}/10^{39}{\rm
erg/s})^{0.19}(\nu_{\rm max}/5{\rm GHz})^{1+\alpha_{\nu}}$, where
$\nu_{\rm max}$ is the high turnover frequency of the synchrotron emission} 
\item{The relation between
$L_{\rm kin}$ and {\it observed} radio core luminosity 
is flattened with respect to the intrinsic one because of the
unaccounted-for Doppler boosting in the small sample at hand. Results
of previous works that have studied the  relationship 
between total extended radio power and kinetic power
\cite{birzan:04,birzan:06}, found both flat slopes and very large
intrinsic scatter (much larger than what we found here). As B{\^ i}rzan et al. (2006)
noticed, such a scatter is "intrinsic to the radio data, for reasons
that include radio aging and adiabatic expansion." This might be
considered as a further argument is support of our approach of using
core radio luminosities, instead, as estimator of the AGN kinetic power.}
\item{The intrinsic variability of the AGN could affect our results
due to the fact that the measured kinetic power is a long-term 
average of the instantaneous power, and that the luminosity of
accreting black holes are log-normally distributed, and therefore
positively skewed. Very little is known about the long-term
(10$^7$-10$^8$ years)
variability properties of low luminosity AGN. 
However, available estimates of rms variability as
a function of luminosity on shorter timescales,
scaling relations with galactic black holes as well as current
estimates of quasar lifetimes all seem to suggest that the variability
cannot be so extreme to affect the results of 
a correlation analysis more dramatically than relativistic beaming does.}  
\end{itemize}
With the aid of these findings, quantitative assessments of
kinetic feedback from supermassive black holes in the radio mode
(i.e. at low dimensionless accretion rates) will be possible based
on accurate determinations of the central engine properties alone, such
as mass, radio core and/or X-ray luminosity. This will provide useful
constraints for AGN feedback models in a cosmological context (Merloni
and Heinz 2007, in prep.).

\section*{Acknowledgments}
We thank the anonymous referee for insightful and constructive
comments. AM thanks Silvia Pellegrini, Debora Sijacki, Martin
Hardcastle and Kazushi Iwasawa for useful discussions. 
SH acknowledges support through NASA grant GO7-8102X.

\appendix
\section{The effect of relativistic beaming on the fundamental plane
relation for active black holes}
\label{app:mc}
We are concerned here with the following question: What is the effect of relativistic beaming on the determination of the
slope of the fundamental plane (FP, MHD03) relation? Or, put it in another way,
can we safely assume that the FP
relation (in any of its incarnations) is free from biases due to
relativistic beaming, and thus suitable to use as a calibrator to
estimate the intrinsic radio core luminosity?

As originally discussed in MHD03, we introduce the ``observed'' FP relation as
\begin{equation}
\label{eq:fp}
\log L_{\rm R,obs}=\xi_{\rm RX} \log L_{\rm
  X}+\xi_{\rm RM} \log M_{\rm BH} + b_{\rm R},
\end{equation}
where the observed correlation coefficients depend, at the 1-sigma level, on
the specific choice of sample, as discussed in KFC06.

The strongest effect on the slope of any intrinsic correlation (i.e. any
correlation between the intrinsic, un-boosted luminosities of jetted
sources) 
comes from the inclusion/exclusion of object with jet axis close to our line of
sight (see e.g. Heinz and Merloni 2004). 
If we define a ``cut-off'' angle, $\theta_c$, such that all
object with inclination $\theta<\theta_c$ are excluded from a
sample, then the slope of the observed correlation deviates from the
intrinsic one as a strong function of $\theta_c$ once the Lorentz
factor $\Gamma >
1/\theta_c$.

In order to assess this effect, we
 have simulated a sample of nearby SMBH according to the selection
 criteria of MHD03 and K\"ording, Falcke and Corbel (2006) [KFC06], 
with a simple Monte Carlo routine. In our fiducial calculation, 
an effective cut-off angle of
$\theta_c=15^{\circ}$ is assumed, 
as these samples have been ``cleaned'' 
of relativistically boosted radio cores by excluding (or correcting
 for) all known Bl Lac objects. 
The radio fluxes are drawn from the observed local (z=0) flat spectrum
radio luminosity function \cite{dunlop:90}. Distances are drawn from a
 distribution that closely resembles that of the soirces in our sample
 (see Table~\ref{tab:table}), and objects with a flux smaller than 0.1 mJy
at 5 GHz and 10$^{-14}$ ergs s$^{-1}$ cm$^{-2}$ in the
 2-10 keV band are excluded. The radio luminosity is then un-boosted
assuming random orientation of the jet between $\theta_c$ and
90$^{\circ}$. The jet Lorentz factor is assumed to be normally
distributed around $\Gamma_{\rm m}$ with variance
$\sigma_{\Gamma}=0.1\Gamma_{\rm m}$, and we studied the effects of
varying $\Gamma_{\rm m}$ between 1.5 and 50.
We fitted the simulated data with the (symmetric) OLS Bisector 
regression algorithm \cite{isobe:90} in order to determine the
relationship between the {\it intrinsic} radio luminosity $L_{\rm R}$ and
the observed one: $\log L_{\rm R}=\xi_{\rm RR}(\Gamma_{\rm m}) \log
 L_{\rm R,obs} + b_{\rm RR}$.
As expected, we found that the larger the mean Lorenz factor of the
 objects, 
the more the observed correlations are skewed away from the $\xi_{\rm RR}=1$
intrinsic slope. In particular, the best fit slopes can be
 approximated by the following log-linear relation:
\begin{equation}
\label{eq:xirr}
\xi_{\rm RR} \simeq 1 - 0.14 \log \Gamma_{\rm m}, 
\end{equation}
that can be regarded as a ``calibrator'' of the fundamental plane
relationship. 

\begin{figure}
\psfig{figure=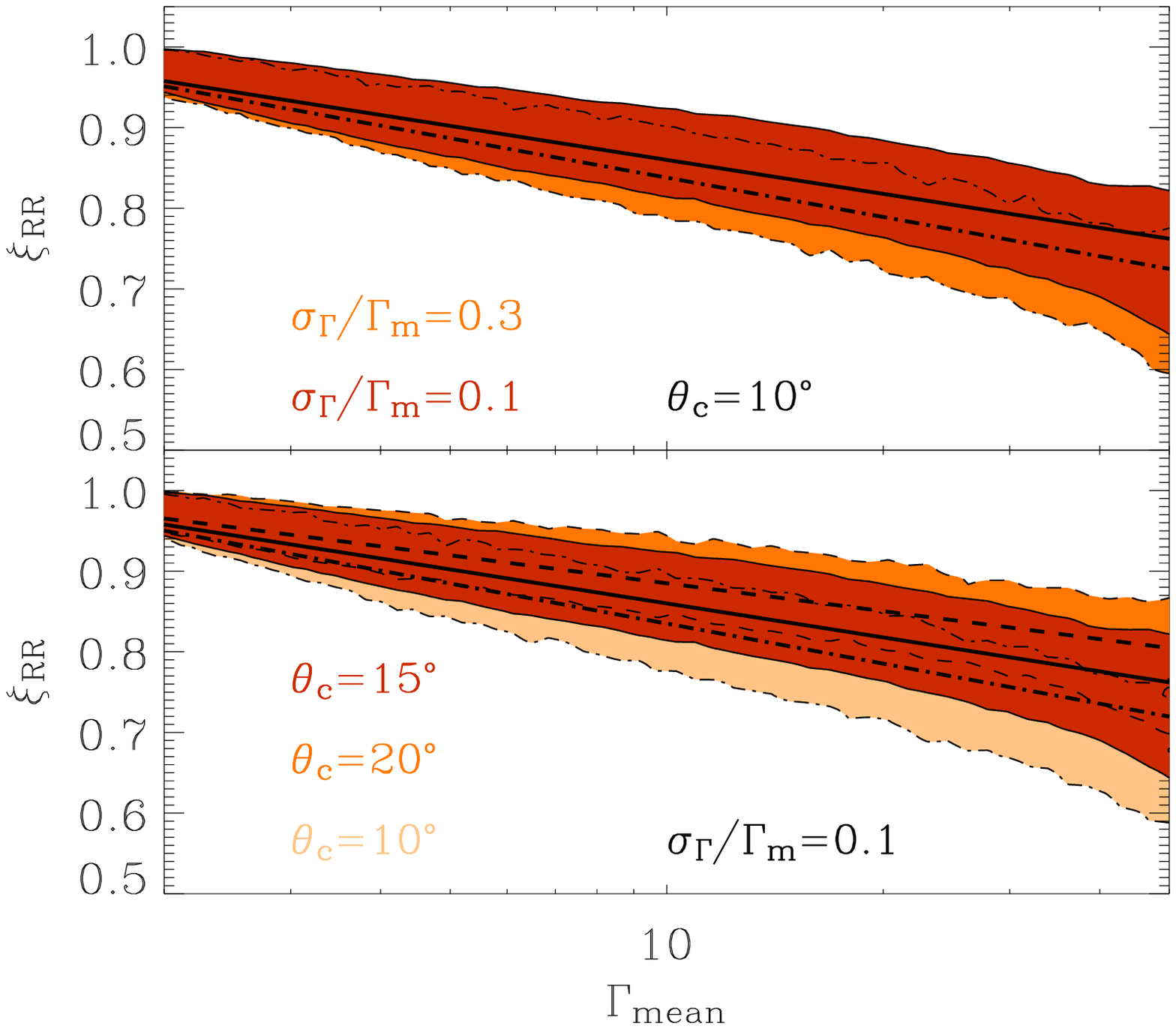,width=0.5\textwidth}
\caption{The effects of relativistic beaming on the observed radio
luminosities in simulated samples resembling those used to derive the
``fundamental plane'' relation. In the upper panel we show, as a
function of the mean jet Lorentz factor, the best-fit slope $\xi_{\rm
RR}$ of the
relation between intrinsic and observed radio core luminosity for two
values of the width of the $\Gamma$ distribution:
$\sigma_{\Gamma}/\Gamma=0.1$ (solid contours and thick solid line) and
$\sigma_{\Gamma}/\Gamma=0.3$ (dot-dashed contours and thick dot-dashed
line). In the lower panel we show $\xi_{\rm RR}$ vs. $\Gamma_{\rm m}$ for three
values of the cut-off angle:
$\theta_c=10^{\circ}$ (dot-dashed contours and thick dot-dashed line),
$\theta_c=15^{\circ}$ (solid contours and thick solid line) and
$\theta_c=20^{\circ}$ (dashed contours and thick dashed line). 
}
\label{fig:fp_gamma}
\end{figure}

The above relation (\ref{eq:xirr}) is plotted in both upper and lower
panels of Fig.~\ref{fig:fp_gamma} as a solid line, while the dark red
shaded areas represent the 1-sigma contours. Figure~\ref{fig:fp_gamma}
also illustrates the effects of slightly varying our fiducial assumptions on
the distribution of the Lorentz factors and on the cut-off angle.
In particular, the upper panel shows the results of increasing the
width of the Lorentz factor distribution by a factor of 3 (from
$0.1\Gamma_{\rm m}$ to $0.3\Gamma_{\rm m}$, dot-dashed lines). In this case the best fit
slope  is approximated by another  log-linear with slope -0.16 (thick
dot-dashed line). This
translates into a difference in the FP slope from our fiducial case around $\Gamma_{\rm m}
\simeq 7$ of just about 3\%, much less than the statistical
uncertainties. Similar is the case when different cut-off angles are
considered. In the lower panel of Fig.~\ref{fig:fp_gamma} we show how
the observed correlations are skewed due to relativistic beaming when
$\theta_c=10^{\circ}$ (dot-dashed contours and thick dot-dashed line)
or $\theta_c=20^{\circ}$ (dashed contours and thick dashed line). Also
in these cases the difference with respect to our fiducial case
corresponds to a difference in the ``corrected'' FP coefficients of
less than $\sim$3\% at  $\Gamma_{\rm m}
\simeq 7$.

In summary, for any value of mean jet Lorentz factor of
the sampled AGN, $\Gamma_{\rm m}$, we have shown that it is possible
to derive statistically an intrinsic (un-boosted)
radio core luminosity based on the observed hard X-ray one, $L_{\rm
X}$ and on the black hole mass, $M_{\rm BH}$ according to:
\begin{eqnarray}
\log L_{\rm R,FP}&=&(1-0.14 \log \Gamma_{\rm m})[\xi_{\rm RX} \log L_{\rm
  X}+\xi_{\rm RM} \log M_{\rm BH} ] \nonumber \\ && + c_{\rm R}(\Gamma_{\rm
  m}), \nonumber
\end{eqnarray}
where we have defined the new constant $c_{\rm R}=\xi_{\rm RR}b_{\rm
R}+b_{\rm RR}$. The term in the first parenthesis in the right hand
side thus represents our simple way to estimate the relativistic
beaming bias introduced in the samples used originally to define the
FP relation. Its numerical value do indeed depends on the assumptions
of our fiducial Monte Carlo model, but to an extent that is negligible
when compared to the statistical uncertainties on the FP parameters themselves.
This above relation, then allows a meaningful statistical test of the
intrinsic correlation between radio core luminosity and kinetic power
of AGN jets, as we show in section~\ref{sec:rad}.

\label{lastpage}


\begin{thebibliography}{}

\bibitem[Aharonian et al. 2004]{aharonian:04}
Aharonian, F., et al., 2004, A\&A, 425, L13

\bibitem[Aitken et al. 2000]{aitken:00}
Aitken, D.K., Greaves, J., Chrysostomou, A., Jenness, T., Holland, W.,
Hough, J. H., Pierce-Price, D., Richer, J., 2000, ApJ, 534, L173 

\bibitem[Akritas \& Siebert 1996]{akritas:96}
{Akritas}, M.~G. and {Siebert}, J., 1996, MNRAS, 278, 919

\bibitem[Allen et al. 2001]{allen:01}
Allen S. W., Fabian A. C., Johnstone R. M., Arnaud K. A. \& Nulsen
P. E. J., 2001, MNRAS, 322, 589

\bibitem[Allen et al. 2006]{allen:06}
{Allen}, S.~W., {Dunn}, R.~J.~H., {Fabian}, A.~C., {Taylor}, G.~B., and
  {Reynolds}, C.~S., 2006, MNRAS, 372, 21

\bibitem[Atoyan \& Dermer 2004]{atoyan:04}
Atoyan, A. \& Dermer, C. D., 2004, ApJ, 617, L123

\bibitem[Baganoff et al. 2001]{baganoff:01}
Baganoff, F. K., et al., 2001, Nature, 413, 45

\bibitem[Baganoff et al. 2003]{baganoff:03}
Baganoff, F. K., et al., 2003, ApJ, 591, 891


\bibitem[Balmaverde, Capetti \& Grandi 2006]{balmaverde:06}
Balmaverde, B., Capetti, A., Grandi, P., 2006, A\&A, 451, 35

\bibitem[Begelman 2001]{begelman:01}
{Begelman}, M.~C., 2001,
\newblock In ASP Conf. Ser. 240: Gas and Galaxy Evolution, {Hibbard}, J.~E.,
  {Rupen}, M., and {van Gorkom}, J.~H., eds.,  363 

\bibitem[Benson et al. 2003]{benson:03}
{Benson}, A.~J., {Bower}, R.~G., {Frenk}, C.~S., {Lacey}, C.~G., {Baugh},
  C.~M., and {Cole}, S., 2003, ApJ, 599, 38

\bibitem[Best et al. 2006]{best:06}
Best, P.N., Kaiser, C.R., Heckman, T.M. \& Kaufmann, G., 2006, MNRAS,
368, L67

\bibitem[Bicknell \& Begelman 1996]{bicknell:96}
Bicknell G. V. \& Begelman M. C., 1996, ApJ, 467, 597

\bibitem[Binney \& Tabor 1995]{binney:95}
{Binney}, J. and {Tabor}, G., 1995, MNRAS, 276, 663

\bibitem[B{\^ i}rzan et al. 2004]{birzan:04}
{B{\^ i}rzan}, L., {Rafferty}, D.~A., {McNamara}, B.~R., {Wise}, M.~W., and
  {Nulsen}, P.~E.~J., 2004, ApJ, 607, 800

\bibitem[B{\^ i}rzan et al. 2006]{birzan:06}
{B{\^ i}rzan}, L., {McNamara}, B.~R., {Carilli}, C.~L., {Nulsen},
P.~E.~J. and {Wise}, M.~W.,
   2006,  to appear in the Proceedings of "Heating vs. Cooling in
Galaxies and Clusters of Galaxies," eds H. B\"ohringer, P. Schuecker,
G. W. Pratt \& A. Finoguenov . astro-ph/0612393

\bibitem[Blandford \& K\"onigl 1979]{blandford:79}
Blandford R. D. \& K\"onigl A., 1979, ApJ, 232, 34


\bibitem[Blandford \& Begelman 1999]{blandford:99}
Blandford R. D. \& Begelman M. C., MNRAS, 303, L1

\bibitem[Blandford \& Begelman 2004]{blandford:04}
Blandford R. D. \& Begelman M. C., MNRAS, 349, 68


\bibitem[Blanton et al. 2003]{blanton:03}
Blanton E. L., Sarazin C. L., Mc Namara B. R., ApJ, 585, 227

\bibitem[B\"ohringer et al. 1993]{boehringer:93}
{B\"ohringer}, H., {Voges}, W., {Fabian}, A.~C., {Edge}, A.~C., and {Neumann},
  D.~M., 1993, MNRAS, 264, L25

\bibitem[Bower et al. 2003]{bower:03}
Bower, G.C., Wright, M. C., Falcke, H., Backer, D.C., 2003, ApJ, 588, 331

\bibitem[Bower et al. 2005]{bower:05}
Bower, G.C., Falcke, H., Wright, M. C., Backer, D.C., 2005, ApJ, 618, L29


\bibitem[Bower et al. 2006]{bower:06}
{Bower}, R.~G., {Benson}, A.~J., {Malbon}, R., {Helly}, J.~C., {Frenk}, C.~S.,
  {Baugh}, C.~M., {Cole}, S., and {Lacey}, C.~G., 2006, MNRAS, 370, 645


\bibitem[Carilli \& Barthel 1996]{carilli:96}
Carilli C. L. \& Barthel P. D., 1996, A\&A Rev., 7, 1 

\bibitem[Cattaneo \& Teyssier 2007]{cattaneo:07}
Cattaneo A. \& Teyssier R., 2007, MNRAS, 376, 1547

\bibitem[Chiaberge, Capetti \& Celotti 1999]{chiaberge:99}
Chiaberge M.,  Capetti A. \& Celotti A., 1999, A\&A, 349, 77


\bibitem[Chiaberge, Capetti \& Macchetto 2005]{chiaberge:05}
Chiaberge M.,  Capetti A. \& Macchetto F. D., 2005, ApJ, 625, 716


\bibitem[Ciotti \& Ostriker 2001]{ciotti:01}
Ciotti L. \& Ostriker J. P., 2001, ApJ, 551, 131

\bibitem[Churazov et al. 2001]{churazov:01}
{Churazov} E., {Br\"uggen} M., Kaiser C. R.,  {B{\" o}hringer} H. \& {Forman} W., 
2001, ApJ, 554, 261

\bibitem[Churazov et al. 2002]{churazov:02}
{Churazov}, E., {Sunyaev}, R., {Forman}, W., and {B{\" o}hringer}, H.,
2002, MNRAS, 332, 729

\bibitem[Churazov et al. 2005]{churazov:05}
{Churazov} E., {Sazonov} S., {Sunyaev} R., {Forman} W., {Jones} C., {B{\" o}hringer}, H.,
2005, MNRAS, 363, L91

\bibitem[Cohen et al. 2007]{cohen:07}
Cohen, M.H., Lister, M.L., Homan, D.C., Kadler, M., Kellerman, K.I.,
Kovalev, Y.Y., Vermeulen, R.C., 2007, ApJ, 658, 232


\bibitem[Croton et al. 2006]{croton:06}
{Croton}, D.~J. et al., 2006, MNRAS, 365, 11


\bibitem[Dalla Vecchia et al. 2004]{dallavecchia:04}
Dalla Vecchia C., Theuns T., Bower R. C., Balogh M. L., Mazzotta P., Frenk
C. S., MNRAS, 2004, 355, 995

\bibitem[De Ruiter et al. 1986]{deruiter:86}
De Ruiter H. R., Parma P., Fanti C., Fanti R., 1986, A\&AS, 65, 111

\bibitem[Di Matteo et al. 2001]{dimatteo:01}
Di Matteo, T., Johnston, R. M., Allen, S. W., \& Fabian, A. C., 2001, ApJL, 550, L19

\bibitem[Di Matteo et al. 2003]{dimatteo:03}
Di Matteo T., Croft R. A. C., Springel V. \& Hernquist L., 2003, ApJ,
593, 56

\bibitem[Done \& Gierli{\'n}ski 2005]{done:05}
Done C. \& Gierli{\'n}ski M., 2005, MNRAS, 364, 208

\bibitem[Dunlop \& Peacock 1990]{dunlop:90}
Dunlop J. S. \& Peacock J. A., 1990, MNRAS, 247, 19

\bibitem[Evans et al. 2006]{evans:06}
Evans D. A., Worrall D. M., Hardcastle M. J., Kraft R. P., Birkinshaw
M., 2006, ApJ, 642, 96

\bibitem[Fabian \& Rees 1995]{fabian:95}
Fabian, A.~C., \& Rees, M. J., 1995, MNRAS, 277, L55

\bibitem[Fabian et al. 2000]{fabian:00}
{Fabian}, A.~C., {Sanders}, J.~S., {Ettori}, S., {Taylor}, G.~B., {Allen},
  S.~W., {Crawford}, C.~S., {Iwasawa}, K., {Johnstone}, R.~M., and {Ogle},
  P.~M., 2000, MNRAS, 318, L65.

\bibitem[Fabian et al. 2001]{fabian:01}
{Fabian}, A.~C., Mushotzky, R.~F., Nulsen, P.~E. J., Peterson, J.~R.,
2001, MNRAS, 321, L20

\bibitem[Fabian et al. 2002]{fabian:02}
Fabian A.~C., Celotti A., Blundell K. M., Kassim N. E., Perley R. A.,
2002, MNRAS, 331, 369

\bibitem[Fabian et al. 2006]{fabian:06}
{Fabian}, A.~C., {Sanders}, J.~S., {Taylor}, G.~B., {Allen},
  S.~W., {Crawford}, C.~S., Johnstone, R.~M., {Iwasawa}, 2006, MNRAS, 366, 417.

\bibitem[Falcke \& Biermann 1996]{falcke:96}
Falcke, H. \& Biermann, P. L., 1996, A\&A, 308, 321

\bibitem[Falcke \& Biermann 1999]{falcke:99}
Falcke, H. \& Biermann, P. L., 1999, A\&A, 342, 49

\bibitem[Falcke \& Markoff 2000]{falcke:00}
Falcke, H. \& Markoff, S., 2000, A\&A, 362, 113

\bibitem[Falcke, K\"ording \& Markoff 2004]{falcke:04}
Falcke H., K\"ording E. \& Markoff S., 2004, A\&A, 414, 895

\bibitem[Ferrarese \& Merritt 2000]{ferrarese:00}
Ferrarese, L. \& Merritt, D., 2000, ApJL, 539, L9


\bibitem[Fender, Gallo \& Jonker 2003]{fender:03}
Fender R. P., Gallo E. \& Jonker P. G., 2003, MNRAS, 343, L99


\bibitem[Forman et al. 2005]{forman:05}
Forman W., et al., 2005, ApJ, 635, 894


\bibitem[Gallo, Fender, \& Pooley 2003]{gallo:03}
Gallo, E., Fender, R. P., \& Pooley, G. G., 2003, MNRAS, 344, 60

\bibitem[Gallo et al. 2005]{gallo:05}
{Gallo}, E., {Fender}, R., {Kaiser}, C., {Russell}, D., {Morganti}, R.,
  {Oosterloo}, T., and {Heinz}, S., 2005, Nature, 436, 819

\bibitem[Gebhardt et al. 2000]{gebhardt:00}
Gebhardt, K. et al., 2000, ApJL, 539, L13


\bibitem[Giovannini et al. 1998]{giovannini:98}
Giovannini, G., Cotton, W. D., Feretti, L., Lara, L., \& Venturi, T.,
1998, ApJ, 493, 632

\bibitem[Hardcastle \& Worrall 2000]{hardcastle:00}
Hardcastle M. \& Worrall D., 2000, MNRAS, 314, 359

\bibitem[Hardcastle, Evans \& Croston 2007]{hardcastle:07}
Hardcastle M., Evans D. \& Croston J., 2007, MNRAS, 376, 1849

\bibitem[Hasinger, Miyaji \& Schmidt 2005]{hasinger:05}
Hasinger G., Miyaji T. \& Schmidt M., 2005, A\&A, 441, 417


\bibitem[Heinz \& Sunyaev 2003]{heinz:03}
Heinz, S. \& Sunyaev, R. A., MNRAS, 343, L59

\bibitem[Heinz 2004]{heinz:04a}
{Heinz}, S., 2004, MNRAS, 355, 835

\bibitem[Heinz \& Merloni 2004]{heinz:04}
{Heinz}, S. \& {Merloni}, A., 2004, MNRAS, 355, L1

\bibitem[Heinz \& Grimm 2005]{heinz:05}
Heinz, S. \& Grimm, H.-J., 2005, ApJ, 633, 384

\bibitem[Heinz, Merloni \& Schwab 2007]{heinz:07}
{Heinz}, S., {Merloni}, A. \& Schwab, J., 2007, ApJL, 658, L9

\bibitem[Ho 2002]{ho:02}
Ho, L. C., 2002, ApJ, 564, 120

\bibitem[Ho 2005]{ho:05}
Ho L. C., 2005, Ap\&SS, 300, 219


\bibitem[Ho \& Ulvestad 2001]{ho:01}
Ho, L. C. \& Ulvestad, J. S., 2001, ApJS, 133, 77


\bibitem[Hopkins, Richards \& Hernquist 2007]{hopkins:07}
Hopkins P. F., Richards G. T., Hernquist L., 2007, ApJ,  654, 731


\bibitem[Isobe et al. 1990]{isobe:90}
Isobe, T., Feigelson, E. D., Akritas, M. and Babu, G. J., 1990, ApJ,
364, 104

\bibitem[Johnstone et al. 2005]{johnstone:05}
Johnstone R. M., Fabian A. C., Morris R. G., Taylor G. B., 2005,
MNRAS, 356, 237

\bibitem[Jones et al. 2007]{jones:07}
Jones C., et al., 2007, to appear in the Proceedings of "Heating vs. Cooling in Galaxies and Clusters of Galaxies," eds H. B\"ohringer, P. Schuecker, G. W. Pratt \& A. Finoguenov  


\bibitem[Kirchbaum et al. 1992]{kirchbaum:92}
Kirchbaum T.P., et al. 1992, A\&A, 260, 33

\bibitem[K{\"o}rding, Falcke \& Corbel 2006]{koerding:06a}
{K{\"o}rding}, E.~G., Falcke, H., and {Corbel}, S., 2006, A\&A, 456,
439 (KFC06)

\bibitem[K{\"o}rding, Fender \& Migliari 2006]{koerding:06b}
{K{\"o}rding} E.~G., {Fender} R. and {Migliari} S., MNRAS, 369, 1451


\bibitem[Kuncic \& Bicknell 2004]{kuncic:04}
Kuncic, Z. \& Bicknell G. V., 2004, ApJ, 616, 669

\bibitem[Kuncic \& Bicknell 2007]{kuncic:07}
Kuncic, Z. \& Bicknell G. V., 2007, To appear in the Proceedings  of
the ``Fifth Stromlo Symposium'', ApSS special issue. arXiv:0705.0791[astro-ph]


\bibitem[Loewenstein et al. 2001]{loewenstein:01}
Loewenstein, M., Mushotzky, R. F., Angelini, L., Arnaud, K. A., \&
Quataert, E., 2001, ApJL, 555, L21


\bibitem[Macchetto et al. 1997]{macchetto:97}
Macchetto F. D., Marconi A., Axon D. J., Capetti A., Sparks W. B.,
Crane P., 1997, ApJ, 489, 579


\bibitem[Maoz 2007]{maoz:07}
Maoz D., 2007, MNRAS, 377, 1696

\bibitem[Marconi \& Hunt 2003]{marconi:03}
Marconi A. \& Hunt L. K., 2003, ApJ, 589, L21

\bibitem[Marconi et al. 2004]{marconi:04}
Marconi A., Risaliti G., Gilli R., Hunt L. K., Maiolino R. \& Salvati
M., 2004, MNRAS, 351, 169

\bibitem[Markowitz \& Edelson 2001]{markowitz:01}
Marcowitz A. \& Edelson R., 2001, ApJ, 547, 684

\bibitem[Martini 2004]{martini:04}
Martini P., 2004, in Carnegie Obs. Astrophys. Ser. 1, Coevolution of
Black Holes and Galaxies, ed. L.C. Ho (Cambridge:CUP), 170

\bibitem[McHardy et al. 2006]{mchardy:06}
{McHardy}, I.~M., {K{\"o}rding}, E., {Knigge}, C., {Uttley}, P., and {Fender},
  R.~P., 2006, Nature, 444, 730


\bibitem[Merloni \& Fabian 2002]{merloni:02}
Merloni A., \& Fabian A.C., 2002, MNRAS, 332, 165 

\bibitem[Merloni, Heinz \& Di Matteo 2003]{merloni:03}
Merloni A., Heinz S. \& Di Matteo T., 2003, MNRAS, 345, 1057 (MHD03)


\bibitem[Merloni \& Heinz 2006]{merloni:06h}
Merloni A. \& Heinz S., 2006, to appear in the Proceedings of the IAU
symposium No. 238 "Black holes: from stars to galaxies - across the
range of masses", Prague, Aug. 2006. V. Karas \& G. Matt
(eds.). astro-ph/0611202 


\bibitem[Merloni 2004]{merloni:04}
Merloni A., 2004, MNRAS, 353, 1053


\bibitem[Merloni et al. 2006]{merloni:06}
Merloni A., K{\"o}rding, E., Heinz S., Markoff, S., Di Matteo T. \&
Falcke, H., 2006, New Astronomy, 11, 567

\bibitem[Muno et al. 2004]{muno:04}
Muno, M. P., Baganoff, F. K., Bautz, M. W., Feigelson E. D., Garmire,
G. P., Morris, M. R., Park, S., Ricker, G. R., Townsley, L. K., 2004,
ApJ, 613, 326

\bibitem[Nagar, Falcke \& Wilson 2005]{nagar:05}
Nagar N. M., Falcke H., Wilson A. S., 2005, A\&A, 435, 521


\bibitem[Nandra et al. 1997]{nandra:97}
Nandra K., George I. M., Mushotsky R. F., Turner T. J., Yaqoob T.,
1997, ApJ, 476, 70


\bibitem[Narayan \& Yi 1995]{narayan:95}
Narayan R. \& Yi I., 1995, ApJL, 452, 710

\bibitem[Nipoti \& Binney 2005]{nipoti:05}
Nipoti C. \& Binney J., 2005, MNRAS, 361, 428

\bibitem[Ogle et al. 1997]{ogle:97}
Ogle P. M., Cohen M. H.,  Miller J. S., Tran H. D., Fosbury R. A. E.,
Goodrich R. W., 1997, ApJ, 482, L37

\bibitem[Panessa et al. 2007]{panessa:07}
Panessa F., Barcons X., Bassani L., Cappi M., Carrera F. J., Ho L. C.,
Pellegrini S., 2007, A\&A, 467, 519

\bibitem[Paolillo et al. 2004]{paolillo:04}
Paolillo M., Schreier E. J., Giacconi R., Koekemoer A. M., Grogin
N. A., 2004, ApJ, 611, 93

\bibitem[Papadakis 2004]{papadakis:04}
Papadakis I. E., 2004, MNRAS, 348, 207

\bibitem[Pellegrini 2005a]{pellegrini:05a}
Pellegrini S., 2005a, ApJ, 624, 155

\bibitem[Pellegrini 2005b]{pellegrini:05b}
Pellegrini S., 2005a, MNRAS, 364, 169


\bibitem[Rafferty et al. 2006]{rafferty:06}
{Rafferty}, D.~A., {McNamara}, B.~R., {Nulsen}, P.~E.~J., and {Wise},
M.~W., 2006, ApJ, 652, 216


\bibitem[Rees et al. 1982]{rees:82}
Rees M. J., Phinney E. S., Begelman M. C., \& Blandford R. D., 1982,
Nature, 295, 17

\bibitem[Reynolds \& Begelman 1997]{reynolds:97}
Reynolds C.~R. \& Begelman M.~C., 1997, ApJ, 487, L135


\bibitem[Sambruna, Eracleous, \& Mushotzky 1999]{sambruna:99}
Sambruna R. M., Eracleous M., \& Mushotzky R. F., 1999, ApJ, 526, 60

\bibitem[Sambruna et al. 2000]{sambruna:00}
Sambruna R. M., Chartas G., Eracleous M., Mushotzky R. F., Nousek
J. A., 2000, ApJ, 532, L91


\bibitem[Shakura \& Sunyaev 1973]{shakura:73}
Shakura, N. I. \& Sunyaev, R. A., 1973, A\&A, 24, 337

\bibitem[Sijacki \& Springel 2006]{sijacki:06}
Sijacki D. \& Springel V., 2006, MNRAS, 366, 397


\bibitem[Simpson et al. 1996]{simpson:96}
Simpson C., Ward M., Clements D. L., Rawlings S., 1996, MNRAS, 281, 509


\bibitem[Soria et al. 2006]{soria:06}
Soria R.


\bibitem[Springel et al. 2005]{springel:05}
{Springel}, V., {Di Matteo}, T., and {Hernquist}, L., 2005, ApJ, 620, L79


\bibitem[Tadhunter et al. 2003]{tadhunter:03}
Tadhunter C., Marconi A., Axon D., Wills K., Robinson T. G., \&
Jackson N., 2003, MNRAS, 342, 861

\bibitem[Taylor et al. 2006]{taylor:06}
Taylor G. B., Sanders J. S., Fabian A. C., Allen S. W., 2006, MNRAS,
365, 705

\bibitem[Terashima et al. 2002]{terashima:02}
Terashima, Y., Iyomoto, N., Ho, L. C., \& Ptak, A. F., 2002, ApJS, 139, 1


\bibitem[Tremaine et al. 2002]{tremaine:02}
Tremaine, S., et al., 2002, ApJ, 574, 554


\bibitem[Uttley, McHardy \& Vaughan 2005]{uttley:05}
Uttley P., McHardy I. M., Vaughan S., 2005, MNRAS, 359, 345

\bibitem[Uttley \& McHardy 2005]{uttley:05b}
Uttley P., McHardy I. M., 2005, MNRAS, 363, 586


\bibitem[Willott et al. 1999]{willott:99}
Willott C. J., Rawlings S., Blundell K. M. \& Lacy M., 1999, MNRAS,
309, 1017


\bibitem[Wu, Yuan \& Cao 2007]{wu:07}
Wu, Q., Yuan, F., Cao, X., 2007, ApJ, in press. arXiv:0706.4124 [astro-ph] 

\bibitem[Young et al. 2002]{young:02}
Young A. J., Wilson A. C., Arnaud K. A., Terashima Y., \& Smith
D. A., 2002, ApJ, 564, 176


\bibitem[Zhao et al. 1993]{zhao:93}
Zhao J.-H., Sumi D. M., Burns J. O., \& Duric N., 1993, ApJ, 416, 51

\bibitem[Zirbel \& Baum 1995]{zirbel:95}
Zirbel E. L. \& Baum S. A., 1995, ApJ, 448, 521

\end{thebibliography}
\end{document}